\documentclass[showpacs,preprintnumbers,amssymb]{revtex4}

\usepackage{graphicx}
\usepackage{epsfig}
\usepackage{dcolumn}
\usepackage{bm}
\usepackage{amsmath}
\usepackage{amssymb}

\def\be{\begin{equation}}
\def\ee{\end{equation}}
\def\bea{\begin{eqnarray}}
\def\eea{\end{eqnarray}}

\def\lsim{\:\raisebox{-0.5ex}{$\stackrel{\textstyle<}{\sim}$}\:}
\def\gsim{\:\raisebox{-0.5ex}{$\stackrel{\textstyle>}{\sim}$}\:}

\def\gsim{\lower0.5ex\hbox{$\:\buildrel >\over\sim\:$}}
\def\lsim{\lower0.5ex\hbox{$\:\buildrel <\over\sim\:$}}

\begin{document}

\preprint{SI-HEP-2011-18}

\title{Muon g-2 and lepton flavor violation in a two Higgs doublets model for the fourth generation}
\author{Shaouly Bar-Shalom}
\email{shaouly@physics.technion.ac.il}
\affiliation{Physics Department, Technion-Institute of Technology, Haifa 32000, Israel}
\author{Soumitra Nandi}
\email{soumitra.nandi@gmail.com}
\affiliation{Physique des Particules, Universit\'e de Montr\'eal, C.P. 6128, succ.\ centre-ville, Montr\'eal, QC, Canada H3C 3J7}
\affiliation{Theoretische Elementarteilchenphysik, Department Physik,
Universit\"at Siegen, D-57068 Siegen, Germany}
\author{Amarjit Soni}
\email{soni@bnl.gov}
\affiliation{Theory Group, Brookhaven National Laboratory, Upton, NY 11973, USA}

\date{\today}

\begin{abstract}
In the minimal Standard Model (SM) with four generations (the so called SM4) and in ``standard"
two Higgs doublets model (2HDM) setups, e.g., the type II 2HDM with four fermion generations,
the contribution of the 4th family
heavy leptons to the muon magnetic moment is suppressed and cannot accommodate the measured $ \sim 3 \sigma$ access
with respect to the SM prediction. We show that in a 2HDM for the 4th generation (the 4G2HDM), which
we view as a low energy effective theory for dynamical electroweak symmetry breaking, with one
of the Higgs doublets
coupling only to the 4th family leptons and quarks (thus effectively addressing their large masses),
the loop exchanges of the
heavy 4th generation neutrino can account for the measured value of the muon anomalous magnetic moment.
We also discuss the sensitivity of the lepton flavor violating decays
$\mu \to e \gamma$ and $\tau \to \mu \gamma$ and of the decay $B_s \to \mu \mu$ to the new couplings
which control the muon g-2 in our model.
\end{abstract}


\maketitle

\section{Introduction}

Particle magnetic moments provide an important and valuable test of QED and of the
Standard Model (SM). In the case of the muon and the electron magnetic moments, both
the experimental measurements and the SM predictions are very precisely known.
However, due to its larger mass, the muon magnetic moment is considered more sensitive to
massive virtual particles and hence to new physics (NP).

In the SM, the total contributions to the muon $g-2$ ($a_{\mu}^{SM}$) can be
divided into three parts: the QED, the electroweak (EW) and the hadronic contributions.
While the QED \cite{qed} and EW \cite{ew} contributions are well understood, the
main theoretical uncertainties lies with the hadronic part which are difficult to control \cite{qcd}.
The hadronic loop contributions cannot be calculated from first principles, so that one relies
on a dispersion relation approach \cite{45844}. At present the available $\sigma(e^+ e^- \to hadrons)$
data are used to calculate the leading-order (LO) and higher-order vacuum polarization contributions to
$a_{\mu}^{SM}$; the estimated contributions are given by \cite{arXiv:0908.4300,hep-ph/0611102}
\be
a^{Had}_{LO} = 6955 (40)(7)\times 10^{-11}, \hskip 10pt a^{Had,Disp}_{NLO} = -98 (1)\times 10^{-11}.
\ee
On the other hand, the hadronic light-by-light contribution cannot be calculated from data,
hence, its evaluation relies on specific models. The latest determination of this term is \cite{nyffeler}
\be
a^{Had}_{lbl} = 116(39)\times 10^{-11}.
\ee
Including all these corrections, the complete SM prediction is given by
\be
a_{\mu}^{SM} = 116591834(2)(41)(26)\times 10^{-11} \label{amuSM}~,
\ee
whereas the current experimentally measured value is \cite{pdg}
\be
a_{\mu}^{exp} = 116592089(54)(33)\times 10^{-11} \label{amuexp}~.
\ee

The SM prediction, therefore, differs from the
the experimentally measured value by (see also \cite{gminus2})
\begin{equation}
a_\mu^{\rm new} = a_\mu^{\rm exp} - a_\mu^{\rm SM} = (255 \pm 80)
\cdot 10^{-11} \label{AMM} ~,
\end{equation}
which allows some room for new physics. For the purpose of this work we are going to assume
that the $\sim 3 \sigma$ discrepancy in Eq.~\ref{AMM} is due to NP, although we are aware that
the estimates of the hadronic contributions have appreciable
uncertainties that may provide part of
the discrepancy.

In most extensions of the SM, new charged or neutral states$^{\footnotemark[1]}$\footnotetext[1]{The new states could be a scalar (S), a pseudoscalar (P), a
vector (V) or an axial-vector (A).},
can contribute to the muon anomalous magnetic moment ($\mu$AMM) at
the one-loop (lowest) level.
For example, the $\mu$AMM plays an important role
in constraining the
supersymmetric (SUSY) parameter space, where, as in the SM,
the leading SUSY contribution to $a_{\mu}$ arises at one-loop, and is found
to be enhanced for large $\tan\beta$. In particular,
as was shown in \cite{muong2-susy}, SUSY can address the observed muon $g-2$ discrepancy for $\tan\beta > 5$ and
$\mu >0$ (Higgsino mass parameter), with typical SUSY masses, of the particles involved in the loops,
in the range $100 ~{\rm GeV} - 500 ~{\rm GeV}$.

Model independent analysis show that (for details see
\cite{gminus2}), for small enough
couplings, scalar exchange diagrams could account for the observed $\mu$AMM
with a scalar mass in the range $480 ~{\rm GeV} - 690 ~{\rm GeV}$,
whereas pseudoscalar and axial-vector one-loop exchanges contribute with the wrong sign
and the one-loop vector exchange contributions are too small.

In this paper we will consider the $\mu$AMM in a new 2HDM framework with
a heavy 4th generation family. Indeed, we will show that the
$\sim 3\sigma$ access (with respect to the SM prediction) shown in Eq.~\ref{AMM} can be
explained by one-loop exchanges of the heavy 4th generation neutrino ($\nu^\prime$)
in a model with two Higgs doublets that we have constructed in \cite{4G2HDM}
and named the 4G2HDM. These new class of two Higgs doublet models
were proposed in \cite{4G2HDM} as viable
low energy effective frameworks for models of 4th generation condensation.
In particular, a theory with new heavy fermionic states
is inevitably cutoff at the near by TeV-scale, where one thus expects
some form of strong dynamics and/or compositeness to occur. Thus, as
was noted already 20 years ago \cite{luty}, the low-energy (i.e., sub-TeV) dynamics of such
a scenario may be more naturally embedded in multi-Higgs theories,
where the new composite scalars are viewed as manifestations
of the several possible bound states of the fundamental heavy fermions.
Besides, our 4G2HDM can naturally (albeit effectively) accommodate
the large (EW-scale) mass of the heavy 4th generation neutrino, which
otherwise remains a cause of concern in theories with a 4th family
of fermions.

We recall that an additional fourth generation of fermions
cannot be ruled out by any symmetry argument, and is not
excluded by EW precision data \cite{ewpt}.
It is also interesting to note, that already the simplest 4th generation extension of the SM, the so called SM4,
has the potential to address some of the current open questions in particle physics, such as the
observed baryon asymmetry \cite{gh08}, the Higgs naturalness problem \cite{Hashimoto:2009ty},
the fermion mass hierarchy
problem \cite{Hung:2009ia} {\rm etc...}.$^{\footnotemark[2]}$\footnotetext[2]{Note that the
neutral Higgs within the SM4 was recently excluded at the LHC
in the range $120~{\rm GeV} \lsim m_{H} \lsim 600~ {\rm GeV}$ \cite{SM4Hbound}.
However, this bound is not relevant to a neutral Higgs of an extended Higgs sector, e.g.,
a 2HDM framework with four generations of
fermions, such as the ones suggested in \cite{4G2HDM,valencia}.}
The SM4 can also accommodate the emerging possible hints for new flavor
physics \cite{SAGMN08,SAGMN10,ajb10B,gh10,Nandi:2010zx,lenz_fourth1}.
However, the SM4 as such cannot explain the observed muon g-2 discrepancy,
see e.g., \cite{hou_muong2}. In fact, even ``standard" 2HDM frameworks
(like the type II 2HDM that underlies the minimal SUSY model)
with an additional 4th generation of heavy fermions, was shown to fail in explaining
the measured $\mu$AMM \cite{hou_muong2}.

In section \ref{sec2} we calculate the $\mu$AMM in the 4G2HDM framework.
In sections \ref{sec3} and \ref{sec4} we consider the constraints on
$\mu$AMM from the lepton flavor violating (LFV) decays
$\mu \to e \gamma$ and $\tau \to \mu \gamma$ and from $B_s \to \mu \mu$,
respectively, and in section \ref{sec5} we summarize our results.

\section{Muon $g-2$ in the 4G2HDM \label{sec2}}

At the tree level the muon magnetic moment is predicted by the Dirac equation
to be $\vec M = g_\mu \frac{e}{2m_\mu} \vec{S}$ with $g_\mu =2$.
The effective vertex of a photon with a charged fermion can in general be written as
\begin{equation}
\bar{u}(p') e \Gamma_\mu u(p) = \bar{u}(p')  e \left[ \gamma_\mu F_1(q^2)
+ \frac{i  \sigma_{\mu\nu} q^\nu}{2m_f} F_2(q^2) \right] u(p) \, ,
\end{equation}
where, to lowest order, $F_1(0) =1$ and $F_2(0) =0$. While $F_1(0)$ remains unity at
all orders due to charge conservation, quantum
corrections yield $F_2(0) \ne 0$. Thus, since
$g_\mu \equiv 2\left(F_1(0) +  F_2(0)\right)$, it follows that $a_\mu \equiv (g_\mu - 2)/2 = F_2(0)$.

\begin{figure}[t]
\vspace{-5mm}
\hspace*{-2mm}  $\gamma$ \hspace{50.5mm} $\gamma$ \\ [10mm]
\hspace*{3mm}
            $\tau'$ \hspace{15mm}   $\tau'$ \hspace{22mm}
              $\phi^{\pm}$ \hspace{15mm}     $\phi^{\pm}$ \\[-30mm]
\vspace{6mm}
\begin{center}
\includegraphics[width=90mm,angle=0]{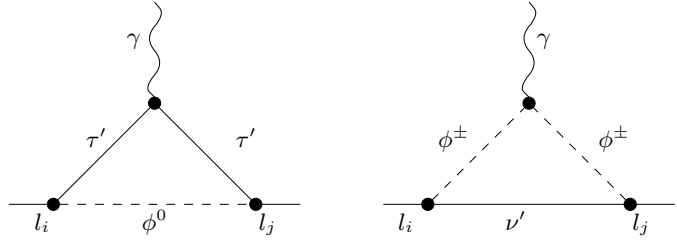}
\end{center}
\vspace{-5mm}
\hspace{0mm}
$l_i$ \hspace{10mm}   $\phi^0$   \hspace{10mm} $l_j$ \hspace{14mm}
$l_i$ \hspace{1cm} $\nu^{\prime}$ \hspace{12mm} $l_j$ \hspace{6.5mm}
\begin{center}
\caption{\sf One-loop diagrams for $l_i \to l_j \gamma$ with charged and neutral scalar exchanges.}
\label{fig:one-loop}
\end{center}
\vspace{-1cm}
\end{figure}
In our 4G2HDM \cite{4G2HDM}  the one-loop contribution to the $\mu$AMM can be subdivided as
\begin{equation}
a_{\mu} = [a_{\mu}]^{SM4}_W + [a_{\mu}]^{4G2HDM}_{\cal H},
\end{equation}
where $[a_{\mu}]^{4G2HDM}_{\cal H}$ contains the charged and neutral Higgs contributions coming from
the one-loop diagrams in Fig.~\ref{fig:one-loop} (see below; the diagrams with
$\tau^\prime$ and $\nu^\prime$ in the loop dominate),
whereas the SM4-like contribution, $[a_{\mu}]^{SM4}_W$, comes from the one-loop
diagram with $W^{\pm} - \nu^\prime$
in the loop and is given by \cite{jlev}
\be
\frac{[a_{\mu}]^{SM4}_W}{|U_{24}|^2} = \frac{G_F m^2_{\mu}}{4\sqrt{2}\pi^2} A(x_{\nu'})~,
\label{ag2sm4}
\ee
where $U_{24}$ is the 24 element of the CKM-like PMNS leptonic matrix,
$x_i = m_i^2/M_W^2$
and the loop function $A(x_i)$ is given by
\be
A(x_i) = \frac{3 x_i^3 \log{x_i}}{(x_i-1)^4} + \frac{4 x_i^3-45 x_i^2 + 33 x_i -10}{6(x_i-1)^3}.
\ee
For values of $m_{\nu'}$ in the range $100 ~{\rm GeV} \lsim m_{\nu^\prime} \lsim 1000~{\rm GeV}$
one finds $1.5\times 10^{-9} \lsim [a_{\mu}]^{SM4}_W/|U_{24}|^2 \lsim 3.0\times 10^{-9}$,
so that for $|U_{24}|^2 << 1$ (as expected) the simple SM4 cannot accommodate the observed discrepancy in $a_\mu$.

Let us recapitulate the salient features of the 4G2HDM setups introduced in  \cite{4G2HDM}. In these models
one of the Higgs fields (the ``heavier'' field) couples only to heavy fermionic states,
while the second Higgs field (the ``lighter'' field)
is responsible for the mass generation of all other (lighter) fermions.
Applying this principle to the 4th generation leptonic sector we
have

\begin{widetext}
\begin{eqnarray}
\mathcal{L}_{Y}= -\bar{E}_{L}
\left( \Phi_{\ell} Y_e \cdot \left( I-{\cal I} \right) +
\Phi_{h} Y_e \cdot {\cal I} \right) e_{R}
-\bar{E}_{L}
\left( \tilde\Phi_{\ell} Y_\nu \cdot \left( I - {\cal I} \right) +
\Phi_{h} Y_\nu \cdot {\cal I} \right)
\nu_{R} + h.c.\mbox{ ,}
\label{eq:LY}
\end{eqnarray}
\end{widetext}
where $f_{L(R)}$
are left(right)-handed fermion fields, $E_{L}$ is the left-handed
$SU(2)$ lepton doublet and $Y_e,Y_\nu$ are general $4\times4$
Yukawa matrices in flavor space. Also, $\Phi_{\ell}$ and $\Phi_{h}$  are the two Higgs doublets, $I$ is the identity matrix and ${\cal I} \equiv {\rm diag}\left(0,0,0,1\right)$. The Yukawa texture of (\ref{eq:LY}) can be realized in terms of a $Z_2$-symmetry under which the fields transform as follows: $\Phi_{\ell}\to-\Phi_{\ell}$, $\Phi_{h} \to \Phi_{h}$, $E_{L}\to E_{L}$, $e_{R} \to- e_{R}$ (for $e=e,\mu,\tau$), $\nu_{R} \to -\nu_{R}$ (for $\nu=\nu_e,\nu_\mu,\nu_\tau$),
and $\tau^\prime_{R}\to \tau^\prime_{R}$,
$\nu^\prime_{R}\to \nu^\prime_{R}$.

From the point of view of the leptonic sector, the Yukawa interaction in (\ref{eq:LY})
is the natural underlying setup that can effectively accommodate the
heavy masses of the 4th generation leptons, by coupling
them to the heavy Higgs doublet. This setup might also be
an effective underlying description of more elaborate constructions in models of warped extra
dimensions \cite{gustavo3}.

The Yukawa interactions between the physical Higgs bosons and the leptonic states are then given by (see \cite{4G2HDM})
\begin{widetext}
\begin{eqnarray}
{\cal L}(h \ell_i \ell_j) &=& \frac{g}{2 m_W} \bar \ell_i \left\{ m_{\ell_i} \frac{s_\alpha}{c_\beta} \delta_{ij}
-  f^h_\beta \cdot
\left[ m_{\ell_i} \Sigma_{ij}^\ell R + m_{\ell_j} \Sigma_{ji}^{\ell \ast} L \right] \right\} \ell_j h \label{Sff1}~, \\
{\cal L}(H \ell_i \ell_j) &=& \frac{g}{2 m_W} \bar \ell_i \left\{ -m_{\ell_i} \frac{c_\alpha}{c_\beta} \delta_{ij}
+ f^H_\beta  \cdot
\left[ m_{\ell_i} \Sigma_{ij}^\ell R + m_{\ell_j} \Sigma_{ji}^{\ell \ast} L \right] \right\} \ell_j H ~, \\
{\cal L}(A \ell_i \ell_j) &=& - i I_\ell \frac{g}{m_W} \bar \ell_i \left\{ m_{\ell_i} \tan\beta \gamma_5 \delta_{ij}
- f_\beta \cdot
\left[ m_{\ell_i} \Sigma_{ij}^\ell R - m_{\ell_j} \Sigma_{ji}^{\ell \ast} L \right] \right\} \ell_j A ~, \\
{\cal L}(H^+ \nu_i e_j) &=& \frac{g}{\sqrt{2} m_W} \bar \nu_i \left\{
\left[ m_{e_j} \tan\beta \cdot U_{ji} - m_{e_k} f_\beta \cdot
U_{ki} \Sigma^{e}_{kj} \right] R \right. \nonumber \\
&& \left. + \left[ -m_{\nu_i} \tan\beta \cdot U_{ji} + m_{\nu_k} f_\beta \cdot
\Sigma^{\nu \ast}_{ki} U_{jk} \right] L
 \right\} e_j H^+ \label{Sff2}~,
\end{eqnarray}
\end{widetext}
with
\begin{eqnarray}
f_\beta \equiv \tan\beta + \cot\beta ~, ~
f^h_\beta \equiv \frac{c_\alpha}{s_\beta} + \frac{s_\alpha}{c_\beta} ~, ~
f^H_\beta &\equiv & \frac{c_\alpha}{c_\beta} - \frac{s_\alpha}{s_\beta} \label{fpar} ~,
\end{eqnarray}
and $\tan\beta$ is the ratio between the two VEVs. Also,
$H^\pm$ is the charged Higgs, $h,H,A$ are the physical neutral Higgs states ($h$ and $H$ are the lighter
and heavier CP-even neutral states, respectively, and $A$ is the neutral CP-odd state),
and $\ell = e$ or $\nu$ with
weak isospin $I_e=-\frac{1}{2}$
and $I_\nu=+\frac{1}{2}$, respectively. Also,
$R(L)=\frac{1}{2}\left(1+(-)\gamma_5\right)$ and $U$ is the $4 \times 4$ leptonic CKM-like
PMNS matrix. Finally,
$\Sigma^e(\Sigma^\nu)$ are new mixing matrices in the charged(neutral)-leptonic sectors, obtained
after diagonalizing the lepton mass matrices
\begin{widetext}
\begin{eqnarray}
\Sigma_{ij}^e = L_{R,4i}^\star L_{R,4j} ~,~
\Sigma_{ij}^\nu = N_{R,4i}^\star N_{R,4j} ~, \label{sigma}
\end{eqnarray}
\end{widetext}
where $L_R,N_R$ are the rotation (unitary) matrices of the right-handed
charged and neutral leptons, respectively. Notice that $\Sigma^e$ and $\Sigma^\nu$ depend
only on the elements of 4th rows of $L_R$ and $N_R$, respectively, which we will
treat as unknowns, i.e., by expressing physical observables
in terms of $N_{R,4i}$ and $L_{R,4i}$ or, equivalently in terms
of $\Sigma_{ij}^e$ and $\Sigma_{ij}^\nu$.$^{\footnotemark[3]}$\footnotetext[3]{Note that
since $N_{R,4i}$ and $L_{R,4j}$ parameterize mixings among the
4th generation and the 1st-3rd generations leptons,  we expect
$\Sigma^\ell_{ij} \ll \Sigma^\ell_{4k}$ for $i,j,k=1,2,3$, see Eq.~\ref{sigma}.}

Following \cite{jlev}, let us redefine the Higgs Yukawa interactions as
\begin{eqnarray}
{\cal L}({\cal H} \ell_i \ell_j) \equiv \bar \ell_i \left[ S^{\cal H}_{\ell_i \ell_j} + P^{\cal H}_{\ell_i \ell_j} \gamma^5 \right]
\ell_j {\cal H}
\label{generic}~,
\end{eqnarray}
with $\ell=e$ or $\nu$ and ${\cal H}=H^+,h,H$ or $A$. Then,
neglecting terms of order
$m_e/m_{\tau^\prime}$ for $e=e,\mu,\tau$ and
terms of order $\Sigma^\ell_{ij}/\Sigma^\ell_{4k}$ for $i,j,k=1,2,3$,
the above scalar and pseudoscalar couplings,  $S^{{\cal H}^0}_{\ell_i \tau^\prime}$,
 $S^{H^-}_{\ell_i \nu^\prime }$ and $P^{{\cal H}^0}_{\ell_i \tau^\prime}$,
 $P^{H^-}_{\ell_i \nu^\prime }$ ($i=1,2$ or 3), which
 mix the 4th generation leptons with the light leptons, are given in our 4G2HDM by
\begin{eqnarray}
&& S^{{\cal H}^0}_{\ell_i \tau^\prime} = - P^{{\cal H}^0}_{\ell_i \tau^\prime} =
\frac{g}{4} \frac{m_\tau^\prime}{m_W} F^{{\cal H}^0} \Sigma_{44}^\nu \delta_{\Sigma_i} ~, \nonumber \\
&& S^{H^-}_{\ell_i \nu^\prime } =  \frac{g}{2 \sqrt{2}} \frac{m_{\tau^\prime}}{m_W} f_\beta U_{44}^\ast
\Sigma_{44}^\nu \left[ \frac{m_{\nu^\prime}}{m_{\tau^\prime}}
\left( 1 - \frac{t_\beta}{f_\beta \Sigma_{44}^\nu} \right) \delta_{U_i} - \delta_{\Sigma_i}  \right] ~, \nonumber \\
&& P^{H^-}_{\ell_i \nu^\prime } =  - \frac{g}{2 \sqrt{2}} \frac{m_{\tau^\prime}}{m_W} f_\beta U_{44}^\ast
\Sigma_{44}^\nu \left[ \frac{m_{\nu^\prime}}{m_{\tau^\prime}}
\left( 1 - \frac{t_\beta}{f_\beta \Sigma_{44}^\nu} \right) \delta_{U_i}  + \delta_{\Sigma_i}  \right]
\label{SPpar}~,
\end{eqnarray}
where ${\cal H}^0=h,H$ or $A$, $F^h=-f_\beta^h$, $F^H=f_\beta^H$, $F^A= i f_\beta$ (see Eq.~\ref{fpar}) and
\begin{eqnarray}
\delta_{U_i} \equiv \frac{U_{i4}^*}{U_{44}^*} ~,~
\delta_{\Sigma_i} \equiv \frac{\Sigma_{4i}^{e *}}{\Sigma_{44}^\nu}
\label{delta12}~,
\end{eqnarray}
which are the small quantities that parameterize the amount of mixing between the 4th generation leptons
and the light leptons of the 1st, 2nd and the 3rd generations. In what follows we will take all
quantities in Eq.~\ref{SPpar} to be real and always
set $U_{44} = \Sigma_{44}^\nu =1$ and $\tan\beta=1$ (for limits on $\tan\beta$ in the
4G2HDM see \cite{4G2HDM}). We note that $a_\mu$ and the branching ratios for the LFV decays
$\ell_i \to \ell_j \gamma$ are proportional to $1+ \cot^2\beta$
(see Eqs.~\ref{dominant} and \ref{dominant2}), so that there is no enhancement
for $\tan\beta >> 1$.

Using Eq.~\ref{SPpar}, the charged and neutral Higgs contributions to $a_{\mu}$ [with
$H^{\pm} - \nu^\prime$ and ${\cal H}^0 - \tau^\prime$ in the loop (${\cal H}^0=h,H$ or $A$), respectively,
see diagrams in Fig.~\ref{fig:one-loop}]
are given by (see also \cite{jlev})
\be
[a_{\mu}]^{4G2HDM}_{H^{\pm}} \approx \frac{m^2_{\mu}}{8 \pi^2}\int^1_0{dx
\frac{x (x-1) \left\{ x \left( \left| S^{H^-}_{\mu \nu^\prime} \right|^2 + \left| P^{H^-}_{\mu \nu^\prime} \right|^2 \right)
+ \frac{m_{\nu^\prime}}{m_{\mu}} \left( \left| S^{H^-}_{\mu \nu^\prime} \right|^2 - \left| P^{H^-}_{\mu \nu^\prime} \right|^2 \right) \right\} }{m^2_{H^-}\, x + m^2_{\nu^\prime}(1-x)}} \label{chargea},
\ee
\be
[a_{\mu}]^{4G2HDM}_{{\cal H}^0} \approx \frac{m^2_{\mu}}{8 \pi^2}\int^1_0{dx
\frac{x^2 \left\{ (1-x) \left( \left| S^{{\cal H}^0}_{\mu \tau^\prime} \right|^2 + \left| P^{{\cal H}^0}_{\mu \tau^\prime} \right|^2 \right)
+ \frac{m_{\tau^\prime}}{m_{\mu}} \left( \left| S^{{\cal H}^0}_{\mu \tau^\prime} \right|^2 - \left| P^{{\cal H}^0}_{\mu \tau^\prime} \right|^2 \right) \right\} }{m^2_{\tau^\prime}\,x + m^2_{{\cal H}^0}(1-x)}}.
\ee

Note that, for the neutral Higgs case, the term proportional to $m_{\nu^\prime}/m_{\mu}$ vanishes
since $|S^{{\cal H}^0}_{\mu \tau^\prime}|=|P^{{\cal H}^0}_{\mu \tau^\prime}|$ (see Eq.~\ref{SPpar}).
Therefore, the dominant contribution, by far, to $a_{\mu}$ comes from the charged Higgs exchange, in particular,
from the second term (proportional to $m_{\nu^\prime}/m_{\mu}$) in the numerator of Eq.~\ref{chargea}, where

\be
\left| S^{H^-}_{\mu \nu^\prime} \right|^2 - \left| P^{H^-}_{\mu \nu^\prime} \right|^2 =
- \frac{g^2}{2} \frac{m_{\nu^\prime} m_{\tau^\prime}}{m_W^2} f_\beta^2 |U_{44}|^2
|\Sigma_{44}^\nu|^2 \cdot {\rm Re} \left\{ \left( 1 - \frac{t_\beta}{f_\beta \Sigma_{44}^\nu} \right) \delta_{U_2} \delta_{\Sigma_2}^\ast \right\}~,
\label{dominant}
\ee
so that $a_\mu$ is proportional to the product $\delta_{\Sigma_2} \cdot \delta_{U_2}$.

\begin{widetext}
\begin{figure}[htb]
\begin{center}
\epsfig{file=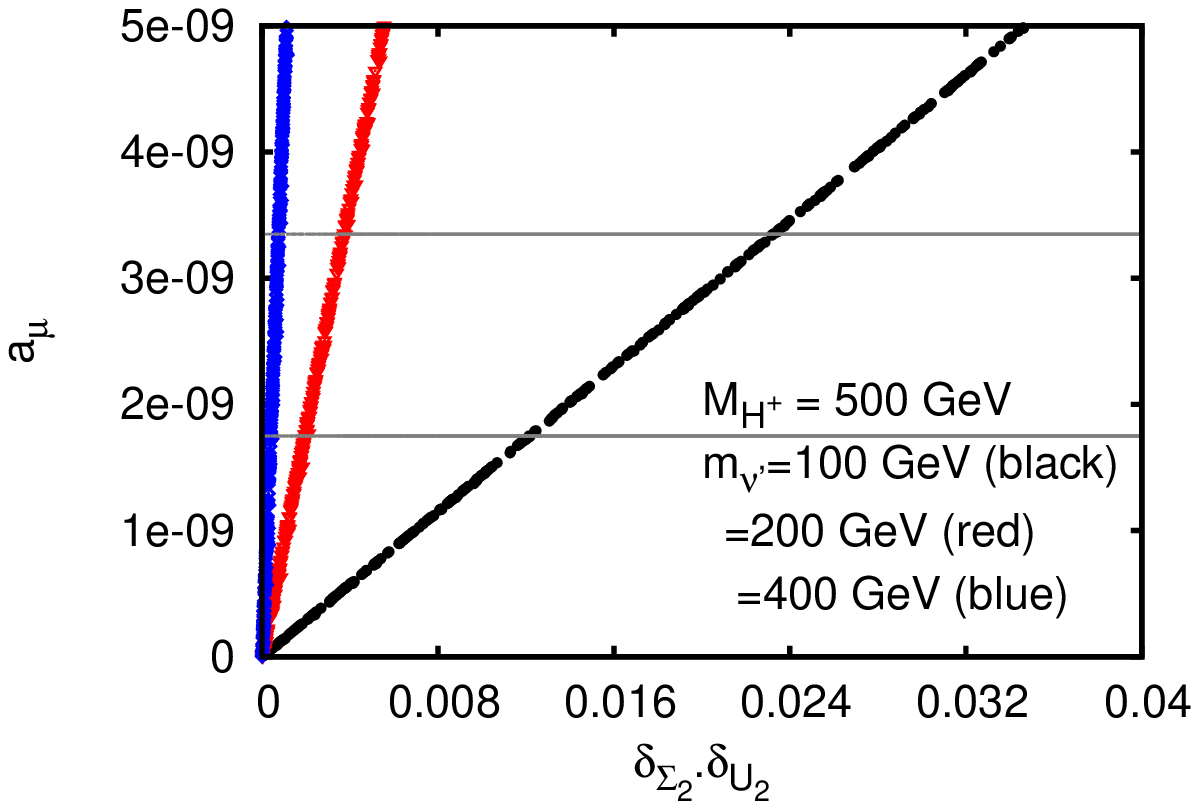,height=7cm,width=7cm,angle=0}
\epsfig{file=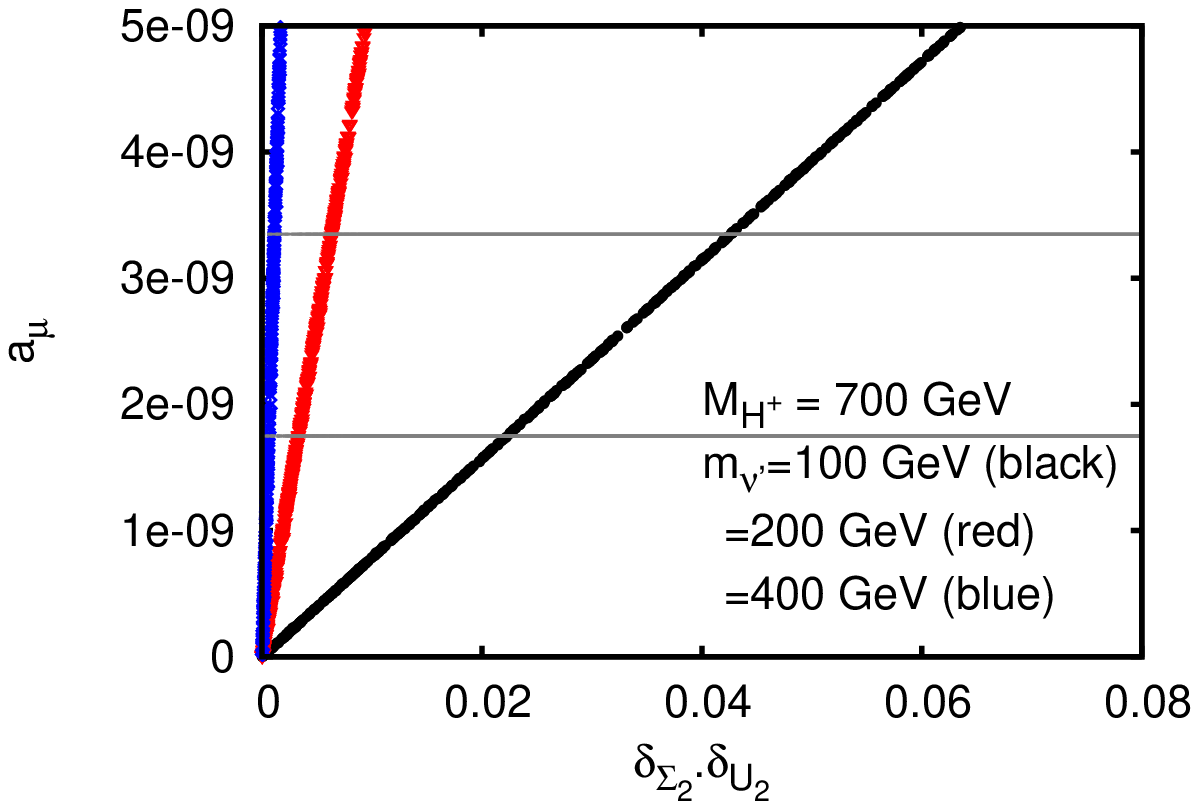,height=7cm,width=7cm,angle=0}
\caption{\emph{The muon $g-2$ as a function of the product
$\delta_{\Sigma_2} \cdot \delta_{U_2}$, for
$m_{\nu^\prime} = 100,~200,~400$ GeV, $m_{\tau^\prime} = m_{\nu^\prime}$ and with $m_{H^+}=500$ GeV (left)
and $m_{H^+}=700$ GeV (right). The horizontal lines are the measured 1-$\sigma$ bounds on $a_\mu$ (see Eq.~\ref{amuexp}).}}
\label{g-2}
\end{center}
\end{figure}
\end{widetext}

In Fig.~\ref{g-2} we plot $a_\mu$ as a function of the product
$\delta_{\Sigma_2} \cdot \delta_{U_2}$ (assuming its real) for several values of
$m_{\nu^\prime}$ and $m_{H^+}$ and fixing
$m_{\tau^\prime} = m_{\nu^\prime}$ ($a_\mu$ depends linearly
on $m_{\tau^\prime}$, see Eq.~\ref{dominant}). Depending on the mass $m_{\nu'}$, we find that
$\delta_{U_2} \cdot \delta_{\Sigma_2} \sim 10^{-3} - 10^{-2}$ is
typically required to accommodate the measured value of $a_\mu$.

In what follows we will consider the constraints from the lepton flavor
violating decays $\ell_i\to \ell_j \gamma$ and from the decay $B_s \to \mu^+ \mu^-$, both
of which are sensitive to the quantities $\delta_{U_2}$ and $\delta_{\Sigma_2}$, as
will be explained below.

\section{Constraints from Lepton flavour violation \label{sec3}}

LFV decays such as $\tau \to \mu\gamma$ and $\mu \to e \gamma$, which are absent in the SM,
are often found useful for constraining
NP models that can potentially contribute to the $\mu$AMM,
as such processes do not suffer from hadronic uncertainties. The current experimental 90\%CL upper
bounds on these LFV decays are \cite{pdg,meg}
\be
Br(\tau \to \mu \gamma) < 4.4 \times 10^{-8} ~~ , ~~ Br(\mu \to e \gamma) < 2.4\times 10^{-12} \label{lfvbounds}~.
\ee

Let us define the amplitude for the transition $\ell_i \to \ell_j \gamma$ as
\begin{equation}
{\cal M}(\ell_i \to \ell_j \gamma) =  \bar{u}_{\ell_j}(p')
\left[i\sigma_{\mu\nu} q^\nu \left( A  +  B \gamma_5 \right) \right] u_{\ell_i}(p)  \epsilon^{\mu\ast} \, ,
\label{liljg}
\end{equation}
where $\epsilon^{\mu\ast}$ is the photon polarization. The decay width is then given by
\be
\Gamma(\ell_i \to \ell_j \gamma) = \frac{m^3_{\ell_i}}{8 \pi}
\left( 1 - \frac{m_{\ell_j}^2}{m_{\ell_i}^2} \right) \left[
\left( 1 + \frac{m_{\ell_j}^2}{m_{\ell_i}^2} \right) \left( |A|^2 + |B|^2 \right)
+4 \frac{m_{\ell_j}}{m_{\ell_i}} \left( |A|^2 - |B|^2 \right) \right] ~.
\ee

Here again, the new 4G2HDM amplitude ${\cal M}(\ell_i \to \ell_j \gamma)^{4G2HDM}$
can be divided as

\be
{\cal M}(\ell_i \to \ell_j \gamma)^{4G2HDM} \equiv {\cal M}^{SM4}_W(\ell_i \to \ell_j \gamma) + {\cal M}^{4G2HDM}_{H^+}(\ell_i \to \ell_j \gamma) + {\cal M}^{4G2HDM}_{{\cal H}^0}(\ell_i \to \ell_j \gamma) ~,
\ee
where ${\cal M}^{SM4}_W(\ell_i \to \ell_j \gamma)$ are the SM4-like W-exchange contribution which is obtained from the diagram (right) of Fig.~\ref{fig:one-loop} with
$\phi^{\pm}$ replaced by $W^{\pm}$ plus the diagrams which contain the self-energy corrections
to the external fermion line $\ell_i$ or $\ell_j$. In particular, using the definition
in Eq.~\ref{liljg} and taking the limit $m_{\ell_j} \to 0$,
the net contribution to ${\cal M}^{SM4}_W(\ell_i \to \ell_j \gamma)$ with
internal $\nu^\prime$ in the loop is given by \cite{beneke_buras}
\be
A^{SM4}_W=B^{SM4}_W = \frac{e G_F m_{\ell_i}}{4 \sqrt{2} \pi^2}  U_{j4} U^{\ast}_{i4} F(x_{\nu^\prime}) ~,
\ee
where $x_i=m^2_i/M^2_W$ and $F(x_i)$ is given by
\be
F(x_i) = \frac{x_i (1-6 x_i + 3 x^2_i + 2 x^3_i - 6 x^2_i \log{x_i})}{4 (1-x_i)^4} ~.
\ee

Here also, we find that ${\cal M}^{SM4}_W(\ell_i \to \ell_j \gamma)$ is much smaller than
the charged and neutral Higgs amplitudes, ${\cal M}^{4G2HDM}_{H^+}(\ell_i \to \ell_j \gamma)$ and ${\cal M}^{4G2HDM}_{{\cal H}^0}(\ell_i \to \ell_j \gamma)$ (calculated from the diagrams in Fig.~\ref{fig:one-loop}),
for which we obtain

\begin{eqnarray}
A_{H^-}^{4G2HDM}  &=& \frac{e}{32 \pi^2} \left\{ (m_{\ell_i}+m_{\ell_j})
\left( {S^{H^-}_{\ell_i \nu^\prime}}^\ast S^{H^-}_{\ell_j \nu^\prime} + {P^{H^-}_{\ell_i \nu^\prime}}^\ast P^{H^-}_{\ell_j \nu^\prime} \right) I_1^{H^+} + 2 m_{\nu^\prime}
\left( {S^{H^-}_{\ell_i \nu^\prime }}^\ast S^{H^-}_{\ell_j \nu^\prime} - {P^{H^-}_{\ell_i \nu^\prime }}^\ast P^{H^-}_{\ell_j \nu^\prime} \right) I_2^{H^+} \right\}~, \label{1stamp}\\
B_{H^-}^{4G2HDM}  &=&\frac{e}{32 \pi^2} \left\{ (m_{\ell_i}-m_{\ell_j})
\left( {S^{H^-}_{\ell_i \nu^\prime}}^\ast P^{H^-}_{\ell_j \nu^\prime} + S^{H^-}_{\ell_j \nu^\prime } {P^{H^-}_{\ell_i \nu^\prime}}^\ast \right) I_1^{H^+} + 2 m_{\nu^\prime}
\left( {S^{H^-}_{\ell_i \nu^\prime}}^\ast P^{H^-}_{\ell_j \nu^\prime} - S^{H^-}_{\ell_j \nu^\prime } {P^{H^-}_{\ell_i \nu^\prime}}^\ast \right)  I_2^{H^+} \right\} ~, \\
A_{{\cal H}^0}^{4G2HDM}  &=& \frac{e}{32 \pi^2} \left\{ (m_{\ell_i}+m_{\ell_j})
\left( {S^{{\cal H}^0}_{\ell_i \tau^\prime}}^\ast S^{{\cal H}^0}_{\ell_j \tau^\prime} + {P^{{\cal H}^0}_{\ell_i \tau^\prime }}^\ast P^{{\cal H}^0}_{\ell_j \tau^\prime} \right) I_1^{{\cal H}^0} + 2 m_{\tau^\prime}
\left( {S^{{\cal H}^0}_{\ell_i \tau^\prime }}^\ast S^{{\cal H}^0}_{\ell_j \tau^\prime} - {P^{{\cal H}^0}_{\ell_i \tau^\prime }}^\ast P^{{\cal H}^0}_{\ell_j \tau^\prime} \right) I_2^{{\cal H}^0} \right\}~, \\
B_{{\cal H}^0}^{4G2HDM}  &=& \frac{e}{32 \pi^2} \left\{ (m_{\ell_i}-m_{\ell_j})
\left( {S^{{\cal H}^0}_{\ell_i \tau^\prime}}^\ast P^{{\cal H}^0}_{\ell_j \tau^\prime} + S^{{\cal H}^0}_{\ell_j \tau^\prime } {P^{{\cal H}^0}_{\ell_i \tau^\prime}}^\ast \right) I_1^{{\cal H}^0} + 2 m_{\tau^\prime}
\left( {S^{{\cal H}^0}_{\ell_i \tau^\prime}}^\ast P^{{\cal H}^0}_{\ell_j \tau^\prime} - S^{{\cal H}^0}_{\ell_j \tau^\prime } {P^{{\cal H}^0}_{\ell_i \tau^\prime}}^\ast \right)  I_2^{{\cal H}^0} \right\} \label{4thamp}~,
\end{eqnarray}
where the loop integrals $I_1^{H^+}$, $I_2^{H^+}$, $I_1^{{\cal H}^0}$ and $I_2^{{\cal H}^0}$ are given by (taking $m_{\ell_i}^2,m_{\ell_j}^2 \ll m_{\cal H}^2,
m_{\nu^\prime}^2,m_{\tau^\prime}^2$):
\bea
I_1^{H^+} &\approx& \int^1_0 dx \frac{x^2(x-1)} {m^2_{H^-}\, x +  m^2_{\nu^\prime}(1-x)} ~, \nonumber \\
I_2^{H^+} &\approx& \int^1_0 dx \frac{x(x-1)}{m^2_{H^-}\, x +  m^2_{\nu^\prime}(1-x)} ~, \nonumber \\
I_1^{{\cal H}^0} &\approx& \int^1_0 dx \frac{x^2(1-x)} {m^2_{\tau^\prime}\, x + m^2_{{\cal H}^0} (1-x)} ~, \nonumber \\
I_2^{{\cal H}^0} &\approx& \int^1_0 dx \frac{x^2}{m^2_{\tau^\prime}\, x + m^2_{{\cal H}^0} (1-x)} ~. \nonumber \\
\eea

The dominant terms in Eqs.~\ref{1stamp}-\ref{4thamp} are the ones proportional to $m_{\nu^\prime}$ from the charged Higgs
exchange contribution,

\begin{eqnarray}
\left( {S^{H^-}_{\ell_i \nu^\prime }}^\ast S^{H^-}_{\ell_j \nu^\prime} - {P^{H^-}_{\ell_i \nu^\prime }}^\ast P^{H^-}_{\ell_j \nu^\prime} \right) &=&
- \frac{g^2}{4} \frac{m_{\nu^\prime} m_{\tau^\prime}}{m_W^2} f_\beta^2 |U_{44}|^2
|\Sigma_{44}^\nu|^2 \left( 1 - \frac{t_\beta}{f_\beta \Sigma_{44}^\nu} \right)
\left( \delta_{U_i}^\ast \delta_{\Sigma_j} + \delta_{U_j} \delta_{\Sigma_i}^\ast \right)
~, \nonumber \\
\left( {S^{H^-}_{\ell_i \nu^\prime }}^\ast P^{H^-}_{\ell_j \nu^\prime} - {S^{H^-}_{\ell_i \nu^\prime }}^\ast P^{H^-}_{\ell_j \nu^\prime} \right) &=&
- \frac{g^2}{4} \frac{m_{\nu^\prime} m_{\tau^\prime}}{m_W^2} f_\beta^2 |U_{44}|^2
|\Sigma_{44}^\nu|^2 \left( 1 - \frac{t_\beta}{f_\beta \Sigma_{44}^\nu} \right)
\left( \delta_{U_i}^\ast \delta_{\Sigma_j} - \delta_{U_j} \delta_{\Sigma_i}^\ast \right)
~, \label{dominant2}
\end{eqnarray}
since the terms proportional to $m_{\tau^\prime}$ in the neutral Higgs exchanges
vanish due to $|S^{{\cal H}^0}_{\ell_i \tau^\prime}|=|P^{{\cal H}^0}_{\ell_i \tau^\prime}|$
(see Eq.~\ref{SPpar}).

We thus find that in our 4G2HDM, the decays $\mu \to e \gamma$ and $\tau \to \mu \gamma$ are sensitive
to $\delta_{U_2}$ and $\delta_{\Sigma_2}$ through the products
$(\delta_{U_2} \delta_{\Sigma_1},\delta_{U_1} \delta_{\Sigma_2})$ and
$(\delta_{U_3} \delta_{\Sigma_2},\delta_{U_2} \delta_{\Sigma_3})$, respectively, so that,
in principle, one can avoid constraints on the quantities $\delta_{U_2}$ and $\delta_{\Sigma_2}$
if $\delta_{U_1},\delta_{U_3},\delta_{\Sigma_1}$ and $\delta_{\Sigma_3}$ are sufficiently small.

\begin{widetext}
\begin{figure}[htb]
\begin{center}
\epsfig{file=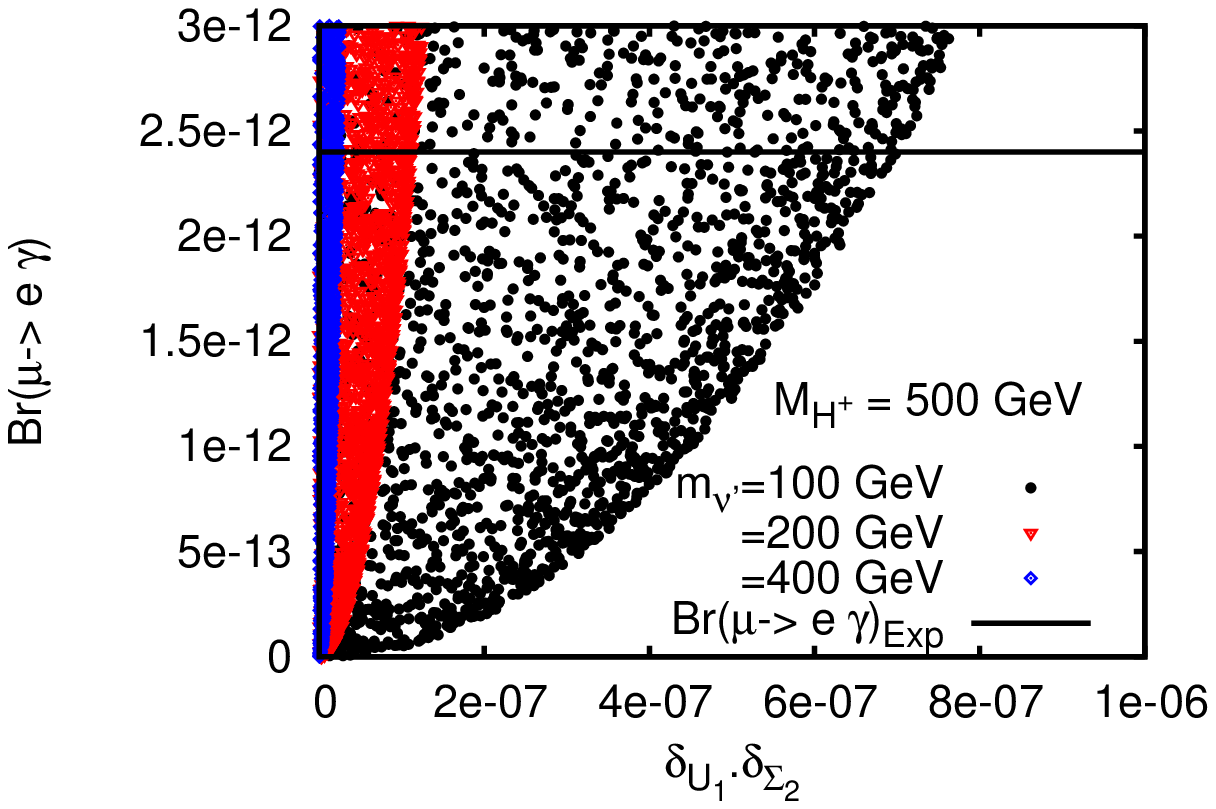,height=7cm,width=7cm,angle=0}
\epsfig{file=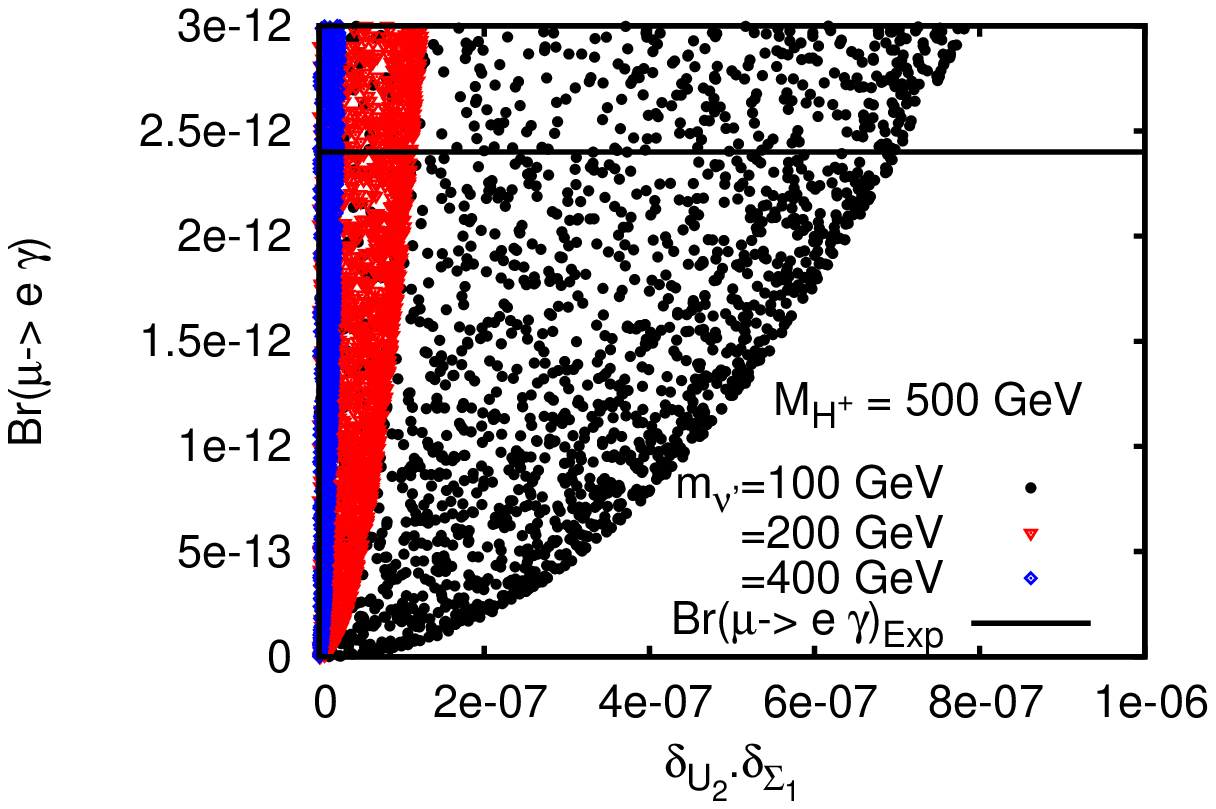,height=7cm,width=7cm,angle=0}
\caption{\emph{$BR(\mu \to e \gamma)$ as a function of
$\delta_{U_1}\cdot \delta_{\Sigma_2}$ (left) and of $\delta_{U_2}\cdot \delta_{\Sigma_1}$ (right), for
$m_{\nu^\prime} = 100,~200,~400$ GeV,
$m_{\tau^\prime} = m_{\nu^\prime}$
and with $m_{H^+}=500$ GeV.
Also shown (horizontal line) is the $90\%CL$ upper limit
on $BR(\mu \to e \gamma)$ (see Eq.~\ref{lfvbounds}).
}}
\label{mueg}
\end{center}
\end{figure}
\end{widetext}

In Figs.~\ref{mueg} we plot $BR(\mu \to e \gamma)$ as a function of $\delta_{U_1} \cdot \delta_{\Sigma_2}$ and
$\delta_{U_2} \cdot \delta_{\Sigma_1}$, for
$m_{\nu^\prime}= 100,~200,~400$ GeV, $m_{H^+}=500$ GeV and fixing
$m_{\tau^\prime} = m_{\nu^\prime}$.
We see that for e.g. $m_{\nu^\prime} = 100$ GeV and for values of $\delta_{U_2}$ and $\delta_{\Sigma_2}$ of
${\cal O}(0.1)$ [for which the product
$\delta_{U_2} \cdot \delta_{\Sigma_2}$ reproduces the measured $a_\mu$
(see Fig.~\ref{g-2})],
$\delta_{U_1}$ and
$\delta_{\Sigma_1}$ are required to be smaller than ${\rm few} \times 10^{-5}$, implying that
$\delta_{U_1} \ll \delta_{U_2}$ and
$\delta_{\Sigma_1} \ll \delta_{\Sigma_2}$.

\begin{widetext}
\begin{figure}[htb]
\begin{center}
\epsfig{file=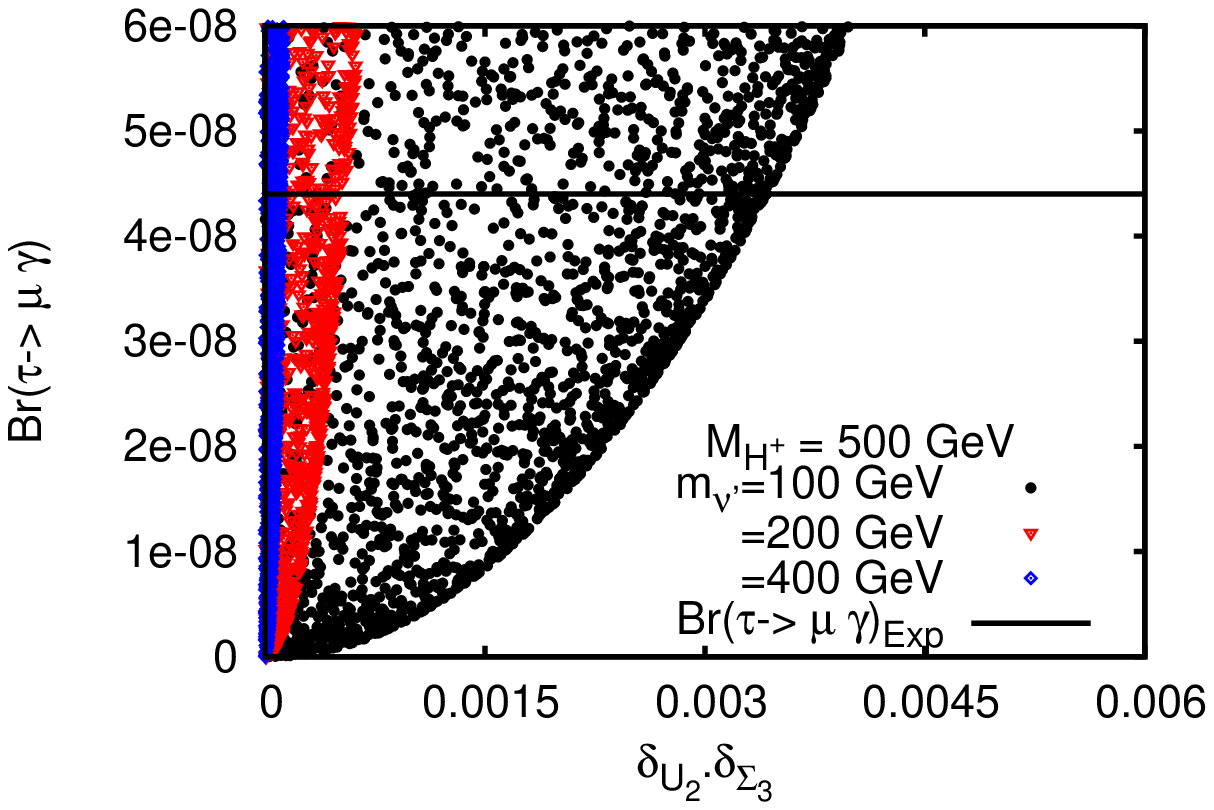,height=7cm,width=7cm,angle=0}
\epsfig{file=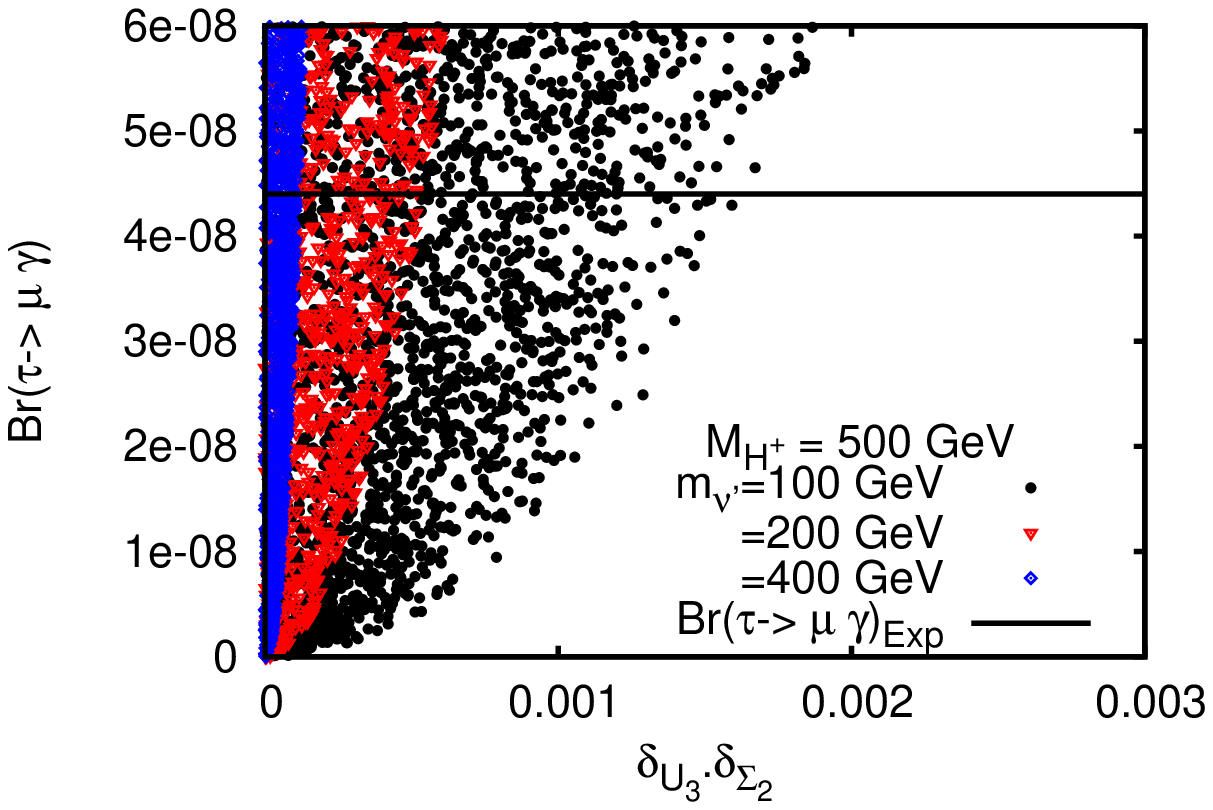,height=7cm,width=7cm,angle=0}
\caption{\emph{$BR(\tau \to \mu \gamma)$ as a function of
$\delta_{U_3}\cdot \delta_{\Sigma_2}$ (left) and $\delta_{U_2}\cdot\delta_{\Sigma_3}$ (right), for
$m_{\nu^\prime} = 100,~200,~400$ GeV, $m_{\tau^\prime} = m_{\nu^\prime}$ and with $m_{H^+}=500$ GeV.
Also shown (horizontal line) is the $90\%CL$ upper limit
on $BR(\tau \to \mu \gamma)$ (see Eq.~\ref{lfvbounds}).
}}
\label{taumug}
\end{center}
\end{figure}
\end{widetext}

\begin{widetext}
\begin{figure}[htb]
\begin{center}
\epsfig{file=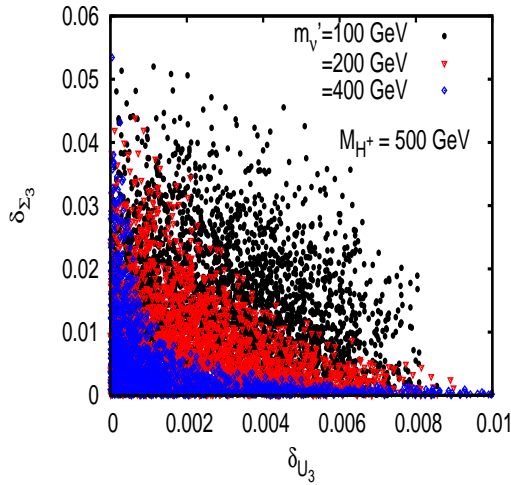,height=7cm,width=7cm,angle=0}
\caption{\emph{Allowed values in the $\delta_{\Sigma_3} -
\delta_{U_3}$ plane, for which
$BR(\tau \to \mu \gamma) < 4.4 \times 10^{-8}$ (i.e., below its 90\%CL bound),
for $m_{\nu^\prime} = 100,~200,~400$ GeV, $m_{\tau^\prime} = m_{\nu^\prime}$ and with $m_{H^+}=500$ GeV.
The couplings $\delta_{U_2}$ and $\delta_{\Sigma_2}$ are chosen
randomly in the range $[0,0.2]$ so that the product
$\delta_{U_2} \cdot \delta_{\Sigma_2}$ reproduces the measured $a_\mu$
(see Fig.~\ref{g-2}).
}}
\label{taumug2}
\end{center}
\end{figure}
\end{widetext}

In Fig.~\ref{taumug} we plot $BR(\tau \to \mu \gamma)$ as a function
of $\delta_{U_2} \cdot \delta_{\Sigma_3}$ and
$\delta_{U_3} \cdot \delta_{\Sigma_2}$, and in
Fig.~\ref{taumug2} we give a scatter plot of the allowed values in the
$\delta_{\Sigma_3} -
\delta_{U_3}$ plane, for which
$BR(\tau \to \mu \gamma) < 4.4 \times 10^{-8}$ (i.e., below its 90\%CL bound).
In both plots
we use
$m_{\nu^\prime}= 100,~200,~400$ GeV, $m_{H^+}=500$ GeV and we fix
$m_{\tau^\prime} = m_{\nu^\prime}$.
The individual couplings $\delta_{U_2}$ and $\delta_{\Sigma_2}$ are
randomly chosen to always be within values that reproduce the measured $a_\mu$
(see Fig.~\ref{g-2}). We see that the products $\delta_{U_2} \cdot \delta_{\Sigma_3}$ and
$\delta_{U_3} \cdot \delta_{\Sigma_2}$ are required to be at most
${\rm few} \times 0.001$, in order to be consistent with the current bounds on
$BR(\tau \to \mu \gamma)$.

We can thus identify a typical benchmark texture for the 4th generation elements
of the CKM-like PMNS matrix, $U_{i4}$, and for the new mixing matrix
$\Sigma^e_{4i}$ that can explain the observed $\mu$AMM and still be consistent with the
current LFV constraints

\begin{eqnarray}
U_{i4} \sim (\Sigma^e_{4i})^T \simeq
\left( \begin{array}{c} \epsilon^5  \\ \epsilon  \\ \epsilon^2 \\ 1
\end{array} \right)
 \label{texture}~,
\end{eqnarray}
where e.g., $\epsilon \sim 0.1$ for $m_{\nu^\prime} = 100$ GeV.

Admittedly, the above texture implies a hierarchical pattern which is
different from the observed hierarchy in the quark's CKM matrix -
usually termed as ``normal". Nonetheless,
without a fundamental theory of flavor,
our insights for flavor should be data driven also in the leptonic sector.
Besides, the above texture is sensitive to the current precision in the measurement
of the muon g-2 which can change e.g., if
more accurate calculations end up showing that part of the hadronic
contributions cannot be ignored.

In Table~\ref{brlfv} we list several representative values
of the couplings $\delta_{U_2}$, $\delta_{\Sigma_2}$ and the products
$(\delta_{U_1},\delta_{\Sigma_2})$,
$(\delta_{U_2},\delta_{\Sigma_1})$, $(\delta_{U_3},\delta_{\Sigma_2})$ and
$(\delta_{U_2},\delta_{\Sigma_3})$ that are consistent with the measured
$\mu$AMM (i.e., $a_{\mu}^{exp}$ given in eq.~\ref{amuexp}), and that give
LFV branching fractions Br$(\tau\to\mu \gamma)$ and Br$(\mu \to e\gamma)$ at the level
of ${\cal O}(10^{-13})$ and ${\cal O}(10^{-9})$, respectively,
that are accessible to near future experiments \cite{meg,superb}.

\begin{table}
\begin{tabular}{|c|c|c|c|c|c|c|c|c|}
\hline
$m_{\nu'}$& $\delta_{U_2} =\delta_{\Sigma_2}$& $a_{\mu}$ & Br$(\mu \to e\gamma)$ &$\delta_{U_1}.\delta_{\Sigma_2}$ & $\delta_{U_2}.\delta_{\Sigma_1}$
& Br$(\tau \to \mu\gamma)$ &$\delta_{U_3}.\delta_{\Sigma_2}$ & $\delta_{U_2}.\delta_{\Sigma_3}$  \\
(GeV) & & $\times 10^9$& $\times 10^{12}$ &$\times 10^7$ & $\times 10^7$ & $\times 10^8$ & $\times 10^4$  & $\times 10^4$ \\
\hline
 &  &  & 1.20 & 1.31 & 4.57 & 1.78 & 1.85 & 20.40  \\
100 &0.12 & 2.07 & 0.64 & 2.45 & 2.25 &0.77  & 3.21  & 5.90   \\
    & & & 0.14 & 1.51 & 0.45 &0.26  & 2.00  & 2.12  \\
\hline
 & &  &1.21 & 0.22 & 0.76 & 2.03 & 0.51 & 3.65  \\
200 & 0.05 & 2.25 &0.62 & 0.51 & 0.25 & 0.79 & 0.41 & 2.25 \\
    & & & 0.11 & 0.43 & 0.61 & 0.18 & 0.16 & 1.07\\
\hline
&  & & 1.39 & 0.16 & 0.04 & 2.14 & 0.73 & 0.12 \\
400 & 0.022 & 2.19 & 0.69 & 0.07 & 0.10 & 0.78 & 0.40 & 0.21\\
    & & & 0.17 & 0.05 & 0.02 & 0.13& 0.11& 0.14 \\
\hline
\end{tabular}
\caption{The calculated $a_\mu$ and the branching fractions for the LFV decays
$\tau\to\mu \gamma$ and $\mu \to e\gamma$ in the 4G2HDM, for several representative values
of the couplings $\delta_{U_2}$, $\delta_{\Sigma_2}$ and the products
$(\delta_{U_1},\delta_{\Sigma_2})$,
$(\delta_{U_2},\delta_{\Sigma_1})$, $(\delta_{U_3},\delta_{\Sigma_2})$ and
$(\delta_{U_2},\delta_{\Sigma_3})$ that are consistent with the measured
$\mu$AMM and which give
Br$(\tau\to\mu \gamma)$ and Br$(\mu \to e\gamma)$ at the level
of ${\cal O}(10^{-13})$ and ${\cal O}(10^{-9})$, respectively.}
\label{brlfv}
\end{table}

\section{Constraints from $B_s \to \ell^+\ell^-$ \label{sec4}}
In the SM, tree level $b\to s $ FCNC transitions are forbidden, and also
the purely leptonic $B_s \to \ell^+ \ell^-$ decays, with
$\ell=e,~ \mu,~\tau$,
suffer from chiral suppression and are therefore very sensitive to new physics.
The SM predicted branching fractions for these decays are
appreciably smaller than those of the semi-leptonic decays. For
example, for $B_s \to \mu^+\mu^-$, the SM prediction is \cite{ajb10B}
\be
Br(B_s \to \mu^+ \mu^-) = (3.2 \pm 0.2) \cdot 10^{-9} ~.
\ee
In the LHC era the current limit on $Br(B_s \to \mu^+ \mu^-)$ has been improved.
A combined analysis by LHCb and CMS, using 0.34$fb^{-1}$ and 1.14$fb^{-1}$ data sample, respectively,
yields \cite{lhcb1}
\be
Br(B_s \to \mu^+ \mu^-)  < 1.08 \times 10^{-8}\;, \hskip 30pt (LHCb+CMS) @ 95\% CL
\ee
whereas the same measurement by CDF-II, using a 7$fb^{-1}$ data sample, gives  \cite{cdfII}
\be
Br(B_s \to \mu^+ \mu^-)  <  4.0 \times 10^{-8} \hskip 35pt CDF @ 95\% CL.
\ee

In fact, LHCb has the sensitivity to measure the $Br(B_s \to \mu^+ \mu^-)$ down to  $ \sim 2 \times 10^{-9}$, which is
about $5\sigma$ smaller than the SM prediction.

\begin{figure}[]
\begin{center}
\resizebox{8cm}{8cm}
{\includegraphics*[53mm,131mm][163mm,244mm]{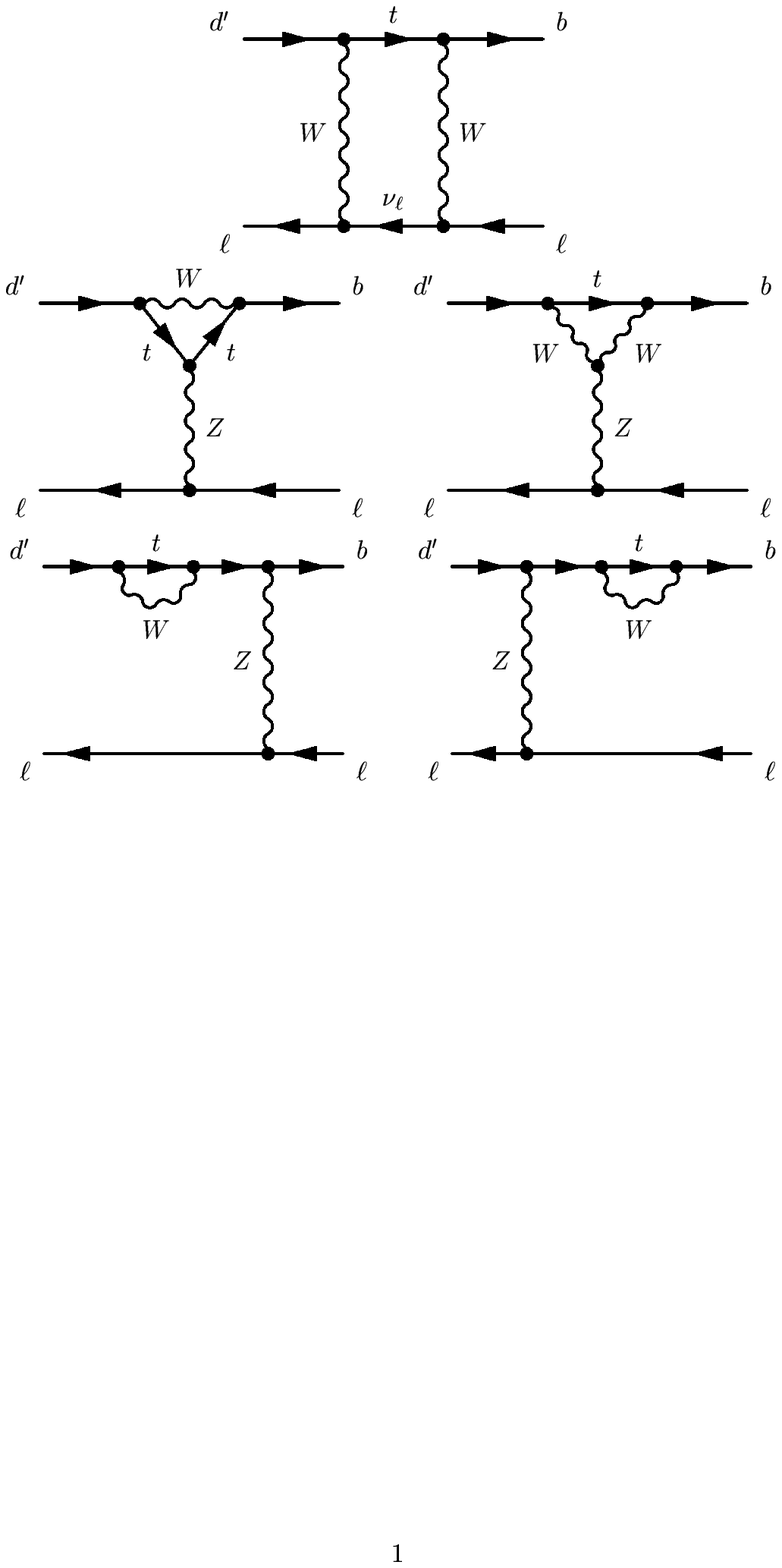}}
\end{center}
\caption{Dominant SM diagrams in the $B_{d'}\to \ell^+\ell^-$ decay.}
\label{smdiags}
\end{figure}
In general, the matrix element for the decay ${\bar B_s}\to \ell^+\ell^-$ can be written as \cite{mBs}
\be
{\cal M} = \frac{G_F \alpha}{2 \sqrt{2} \pi \sin\theta_W^2} \left[ F_{S}\,{\bar \ell} \ell + F_{P}\,{\bar \ell}\gamma_5 \ell
+  F_{A}\,P^{\mu} {\bar \ell} \gamma_{\mu}\gamma_5 \ell \right], \,
\label{ampbsll}
\ee
where $P^{\mu}$ is the four momentum of the initial $B_s$ meson and $F_{i}$'s are functions of Lorentz invariant quantities.
Squaring the matrix and summing over the lepton spins, we obtain the branching fraction
\be
Br({\bar B_s}\to \ell^+ \ell^-) = \frac{G_F^2 \alpha^2 M_{B_s} \tau_{B_s}}{64 \pi^3} \sqrt{1- \frac{4 m_{\ell}^2}{M_{B_s}^2}}
\left[\left(1- \frac{4 m_{\ell}^2}{M_{B_s}^2}\right) |F_S|^2 + |F_P + 2 m_{\ell} F_A|^2\right].
\ee

In the SM, the dominant effect in ${\bar B_s}\to \ell^+ \ell^-$ arise from the diagrams shown in
Fig.~\ref{smdiags}, which contribute to $F_A$ in Eq.~\ref{ampbsll}. At next-to-leading (NLO) QCD corrections,
the net contribution
in $F_A$ is given by \cite{Buchalla:1993bv,Misiak:1999yg}
\be
F_A^{SM} = -\,i\,f_{B_s}\,V_{tb} V_{ts}^{\ast}\, Y(x_t) = -\,i\,f_{B_s}\,V_{tb} V_{ts}^{\ast} \times 0.997 \left[\frac{m_t(m_t)}{166 {\rm GeV}}\right]^{1.55}.
\ee

In the SM4 it is again $F_A$ which receives a non-zero contribution:
\be
F_A^{SM4} = (F_A^{SM4})^{peng} + (F_A^{SM4})^{box},
\ee
where $(F_A^{SM4})^{peng}$ is the contributions from the Z-penguin diagrams in
Fig.~\ref{smdiags} with the top quark replaced by a $t^\prime$, and $(F_A^{SM4})^{box}$ is the contribution
from the box diagram in Fig.~\ref{smdiags} with the replacement $t \to t'$ and $\nu_{\ell}\to \nu'$.
At leading-order (LO) in QCD, they are given by
\be
(F_A^{SM4})^{peng} = -\, i\,f_{B_s} V_{t'b} V_{t's}^{\ast} \frac{x_{t'}}{8}\left[\frac{6-x_{t'}}{1-x_{t'}} + \frac{3 x_{t'} + 2}{(1-x_{t'})^2} \ln{x_{t'}}\right],
\ee
\be
(F_A^{SM4})^{box} = -\, i\,f_{B_s}\,V_{t'b} V_{t's}^{\ast}\,|U_{24}|^2 \frac{B_0(x_{t'},x_{\nu^\prime})}{4},
\ee
with
\bea
B_0(x_i,x_j) &=& x_i\,x_j\left[- \frac{3}{4}\frac{1}{(1-x_i)(1-x_j)} + \frac{\ln{x_i}}{(x_i-x_j)(1-x_i)^2}
\left(1-2 x_i + \frac{x_i^2}{4}\right) \right. \nonumber \\
&{}& \hskip 15pt \left.  +  \frac{\ln{x_j}}{(x_j-x_i)(1-x_j)^2} \left(1-2 x_j + \frac{x_j^2}{4}\right) \right].
\eea
In our 4G2HDM, we have additional contributions to $F_A$ coming from the charged Higgs exchange penguin and box
diagrams (replacing $W^+ \to H^+$ in Fig.~\ref{smdiags}) and, in addition, there are new
contributions to $F_S$ and $F_P$. For our purpose, we are interested
only in the diagrams that are sensitive to the $\ell^{\pm} \nu' H^{\pm}$ vertex, which are, therefore,
directly related to
the muon g-2. The dominant diagrams that contribute to the ${\bar B_s}\to \ell^+ \ell^-$ decay with this vertex are the
Higgs-exchange box diagrams in Fig.~\ref{smdiags}, where one or two $W$-bosons are replaced by $H^+$ and
$(t, \nu_{\ell})$ are being replaced by both $(t, \nu^\prime)$ and $(t^\prime, \nu^\prime)$.
Thus, the net contributions to $F_S$, $F_P$ and $F_A$ in Eq.~\ref{ampbsll} can be written as
\be
F_S = F_S^{H}, \hskip 10pt F_P = F_P^{H} \hskip 10pt and \hskip 10pt F_A = F_A^{SM} + F_A^{SM4} + F_A^{H}~,
\ee
where $F_S^{H},~F_P^{H}$ and $F_A^{H}$ are the contributions from the dominant new Higgs-exchange box diagrams,
which, for the $(t^\prime, \nu^\prime)$ exchange, are given by
\bea
F_S^{H} &=& (F_S)_{HH}^{box} + (F_S)_{WH}^{box}   ,\nonumber \\
 & = & - \frac{i\,2 f_{B_s}}{ g^2}\,\frac{M_{B_s}^2}{m_b + m_s}\left[ \frac{m_{\nu'} m_{t'}}{M_{H^-}^4}
 \left({S_{t's}^{{H^-}}}^{\ast} P_{t'b}^{H^-} -
S_{t'b}^{H^-} {P_{t's}^{{H^-}}}^{\ast}\right)
\left( \left| S^{H^-}_{\mu \nu^\prime} \right|^2 - \left| P^{H^-}_{\mu \nu^\prime} \right|^2 \right) B_{HH}^S(y_{t'},y_{\nu'})\right. \nonumber \\
&{}& \hskip 2.7cm  + \frac{g}{2 \sqrt{2} m_W}\left\{V_{t'b} U_{24}^{\ast}\left({S_{t's}^{H^-}}^{\ast}  + {P_{t's}^{\cal H}}^{\ast}\right) \left({S_{\mu\nu'}^{H^-}}  +
{P_{\mu\nu'}^{H^-}}\right) - V_{t's}^{\ast} U_{24}\left({S_{t'b}^{H^-}}  + {P_{t'b}^{H^-}}\right) \right. \nonumber \\
&{}&\hskip 2.7cm \left.\left. \left({S_{\mu\nu'}^{H^-}}^{\ast}  +
{P_{\mu\nu'}^{H^-}}^{\ast}\right)\right\} B_{WH}^S(x_{t'},x_{\nu'},x_H) \right]
\label{FSH}
\eea
\bea
F_P^H &=& (F_P)_{HH}^{box} + (F_P)_{WH}^{box} ,\nonumber \\
 & = & -\frac{i\,2 f_{B_s}}{ g^2}\,\frac{M_{B_s}^2}{m_b + m_s}\left[\frac{m_{\nu'} m_{t'}}{M_{H^-}^4}
 \left({S_{t's}^{H^-}}^{\ast} P_{t'b}^{H^-}
- S_{t'b}^{H^-} {P_{t's}^{H^-}}^{\ast}\right)
\left(P_{\mu\nu'}^{H^-} {S_{\mu\nu'}^{H^-}}^{\ast} - S_{\mu\nu'}^{H^-} {P_{\mu\nu'}^{H^-}}^{\ast}\right) B_{HH}^P(y_{t'},y_{\nu'})\right. \nonumber \\
&{}& \hskip 2.7cm  + \frac{g}{2 \sqrt{2} m_W} \left\{ V_{t'b} U_{24}^{\ast} \left({S_{t's}^{H^-}}^{\ast}  +
{P_{t's}^{H^-}}^{\ast}\right) \left({S_{\mu\nu'}^{H^-}}  +
{P_{\mu\nu'}^{H^-}}\right) + V_{t's}^{\ast} U_{24}\left({S_{t'b}^{H^-}}  + {P_{t'b}^{H^-}}\right) \right. \nonumber \\
&{}&\hskip 2.7cm \left.\left. \left({S_{\mu\nu'}^{H^-}}^{\ast}  +
{P_{\mu\nu'}^{H^-}}^{\ast}\right)\right\} B_{WH}^P(x_{t'},x_{\nu'},x_H) \right]
\label{FPH}
\eea
\bea
F_A^H &=& (F_A)_{HH}^{box} + (F_A)_{WH}^{box} ,\nonumber \\
 & = &- \frac{i\,2 f_{B_s}}{ g^2}\,\left[ \frac{1}{4\,M_{H^-}^2} \left(S_{t's}^{H^-} {P_{t'b}^{H^-}}^{\ast} +
 {S_{t'b}^{H^-}}^{\ast}
P_{t's}^{H^-}\right) \left(P_{\mu\nu'}^{H^-} {S_{\mu\nu'}^{H^-}}^{\ast} +
S_{\mu\nu'}^{H^-} {P_{\mu\nu'}^{H^-}}^{\ast}
\right) B_{HH}^A(y_{t'},y_{\nu'})\right. \nonumber \\
&{}& \hskip 1.2cm  + \frac{g}{2 \sqrt{2} m_W} \frac{m_{\nu'} m_{t'}}{m_W^2} \left\{ V_{t'b} U_{24}^{\ast} \left({P_{t's}^{H^-}}^{\ast}  - {S_{t's}^{H^-}}^{\ast}\right)
\left({S_{\mu\nu'}^{H^-}} - {P_{\mu\nu'}^{H^-}}\right) + V_{t's}^{\ast} U_{24} \left({P_{t'b}^{H^-}}  -
{S_{t'b}^{H^-}}\right)  \right. \nonumber \\
&{}& \hskip 1.2cm \left.\left. \left({S_{\mu\nu'}^{H^-}}^{\ast} - {P_{\mu\nu'}^{H^-}}^{\ast}\right)\right\} B_{WH}^A(x_{t'},x_{\nu'},x_H) \right],
\label{formfactors}
\eea
where $y_i = m_i^2/M_{H^-}^2$, $x_i = m_i^2/M_W^2$ and the loop-functions are given by
\bea
B_{HH}^S(y_i,y_j) &=& B_{HH}^P(y_i,y_j) =
\frac{1}{y_i-y_j}\left\{\frac{y_j \ln{y_j}}{(1-y_j)^2} - \frac{y_i \ln{y_i}}{(1-y_i)^2}\right\}-\frac{1}{(1-y_i)(1-y_j)} \nonumber \\
B_{WH}^S(x_i,x_j,x_H) &=& B_{WH}^P(x_i,x_j,x_H) =  \frac{x_H^2 \ln{x_H}}{(x_H-1)(x_H-x_i)(x_H-x_j)} \nonumber \\
&{}& + \frac{x_i^2 \ln{x_i}}{(x_i-1)(x_i-x_H)(x_i-x_j)} + \frac{x_j^2 \ln{x_j}}{(x_j-1)(x_j-x_H)(x_j-x_i)} \nonumber \\
B_{HH}^A(y_i,y_j) &=& \frac{y_j^2 \ln{y_j}}{(1-y_j)^2\,(y_j-y_i)} + \frac{y_i^2 \ln{y_i}}{(1-y_i)^2 \,(y_i-y_j)} +\frac{1}{(1-y_i)(1-y_j)} \nonumber \\
B_{WH}^A(x_i,x_j,x_H) &=&  \frac{\,x_H \ln{x_H}}{(x_H-1)(x_H-x_i)(x_H-x_j)} + \frac{x_i \ln{x_i}}{(x_i-1)(x_i-x_H)(x_i-x_j)} \nonumber \\
&{}&     + \frac{x_j \ln{x_j}}{(x_j-1)(x_j-x_H)(x_j-x_i)} ~.
\eea
Also, $S_{\mu \nu^\prime}^{H^-}$, $P_{\mu \nu^\prime}^{H^-}$ are defined in Eq.~\ref{SPpar} and
\bea
S_{t'b}^{H^-} &=&
 (m_b - m_{t'}) t_{\beta} V_{t'b} + f_{\beta} V_{t'b} \left[ \left(m_{t'} \Sigma_{44}^{u*} -
 m_b \Sigma_{33}^d \right) +
\left(m_t \Sigma_{34}^{u*} \frac{V_{tb}}{V_{t'b}} - m_{b'} \Sigma_{43}^d \frac{V_{t'b'}}{V_{t'b}}\right)\right] , \nonumber \\
P_{t'b}^{H^-} &=&   (m_b + m_{t'}) t_{\beta} V_{t'b} - f_{\beta} V_{t'b}
\left[ \left(m_{t'} \Sigma_{44}^{u*} +m_b \Sigma_{33}^d \right) +
\left(m_t \Sigma_{34}^{u*} \frac{V_{tb}}{V_{t'b}} + m_{b'} \Sigma_{43}^d \frac{V_{t'b'}}{V_{t'b}}\right)\right] , \nonumber \\
S_{t's}^{H^-} &=& - P_{t's}^{H^-} =  -  m_{t'} t_{\beta} V_{t's}+ f_{\beta}
\left(m_{t'} V_{t's} \Sigma_{44}^{u*} + m_t V_{ts} \Sigma_{34}^{u*} \right) ~.
 \label{qcoup}
\eea

For the couplings $S_{t'b}^{H^-},~S_{t's}^{H^-},~P_{t'b}^{H^-},~P_{t's}^{H^-}$
we use the 4G2HDM Yukawa terms in the quark sector as given in \cite{4G2HDM}, where
$\Sigma^u$ and $\Sigma^d$ are the corresponding new mixing matrices in the up and down-quark sectors,
respectively, obtained
after diagonalizing the quarks mass matrices. In particular, adopting the type I
4G2HDM of \cite{4G2HDM}, these matrices are given by
\begin{eqnarray}
\Sigma_{ij}^d = D_{R,4i}^\star D_{R,4j}~, ~
\Sigma_{ij}^u = U_{R,4i}^\star U_{R,4j} ~, \label{sigmaud}
\end{eqnarray}
in analogy with Eq.~\ref{sigma}, where $D_R,U_R$ are the rotation (unitary) matrices of the right-handed
down and up-quarks, respectively. They can be approximated by (see \cite{4G2HDM})

\begin{eqnarray}
\Sigma^d \simeq \left(\begin{array}{cccc}
0 & 0 & 0 & 0 \\
0 & 0 & 0 & 0 \\
0 & 0 &  |\epsilon_b|^2 & \epsilon_b^\star \\
0 & 0 &  \epsilon_b  & \left( 1- \frac{|\epsilon_b|^2}{2} \right)
\end{array}\right)~,~
\Sigma^u \simeq \left(\begin{array}{cccc}
0 & 0 & 0 & 0 \\
0 & 0 & 0 & 0 \\
0 & 0 &  |\epsilon_t|^2 & \epsilon_t^\star \\
0 & 0 &  \epsilon_t  & \left( 1- \frac{|\epsilon_t|^2}{2} \right)
\end{array}\right)
 \label{sigmaI}~,
\end{eqnarray}
so that $\Sigma^{u,d} = 0$ if $i ~ {\rm or} ~ j \neq 3,4$, and the $34$ blocks are parameterized
by the quantities $\epsilon_b$ and $\epsilon_t$. As in \cite{4G2HDM},
a natural choice that we will adopt below is $\epsilon_b \sim m_b/m_{b^\prime} << \epsilon_t$.
Thus, neglecting terms of ${\cal O}(m_b/m_t)$ (and, therefore, neglecting also terms proportional
to $\epsilon_b$), the couplings $S_{t'b}^{H^-},~S_{t's}^{H^-},~P_{t'b}^{H^-},~P_{t's}^{H^-}$
  can be approximated by
 \bea
S_{t'b}^{H^-} &=& - P_{t'b}^{H^-} \approx  \frac{m_{t'}}{t_{\beta}} V_{t'b} +  m_t f_\beta \epsilon_t V_{tb} ~,
\nonumber \\
S_{t's}^{H^-} &=& - P_{t's}^{H^-} \approx
 \frac{m_{t'}}{t_{\beta}} V_{t's} +  m_t f_\beta \epsilon_t V_{ts}  ~,
\label{qcoup2}
\eea
leading to $F_S^{H} \to 0$, $F_P^H \to 0$ and
\bea
F_A^H &\approx&  \frac{i\,2 f_{B_s}}{ g^2}\,\left[ \frac{S_{t's}^{H^-} {S_{t'b}^{H^-}}^{\ast}}{M_{H^-}^2}
{\rm Re} \left( S_{\mu\nu'}^{H^-} {P_{\mu\nu'}^{H^-}}^{\ast} \right)
B_{HH}^A(y_{t'},y_{\nu'})\right. \nonumber \\
&{}&  + \frac{g}{\sqrt{2} m_W} \frac{m_{\nu'} m_{t'}}{m_W^2} \left\{ V_{t'b} U_{24}^{\ast} {S_{t's}^{H^-}}^{\ast}
\left({S_{\mu\nu'}^{H^-}} - {P_{\mu\nu'}^{H^-}}\right) + V_{t's}^{\ast} U_{24} {S_{t'b}^{H^-}}
 \left. \left({S_{\mu\nu'}^{H^-}}^{\ast} - {P_{\mu\nu'}^{H^-}}^{\ast}\right)\right\} B_{WH}^A(x_{t'},x_{\nu'},x_H) \right],
\label{formfactors2}
\eea

Notice that the term $U_{24}^\ast \cdot \left( {S_{\mu\nu'}^{H^-}} - {P_{\mu\nu'}^{H^-}} \right)$ is proportional
to $\left(\delta_{U_2} \right)^2$ and ${\rm Re} \left( S_{\mu\nu'}^{H^-} {P_{\mu\nu'}^{H^-}}^{\ast} \right)$
is proportional to both $\left(\delta_{U_2} \right)^2$ and $\left(\delta_{\Sigma_2} \right)^2$.
Thus, there is a net effect in $B_s \to \mu \mu$ from the charged-Higgs box diagrams even
when one of the small quantities that control the muon g-2 vanishes, i.e., when either $\delta_{U_2} \to 0$ or
$\delta_{\Sigma_2} \to 0$, for which cases
$[a_{\mu}]^{4G2HDM}_{H^{\pm}} \to 0$ (see Eq.~\ref{chargea}). For example, for $\delta_{U_2} \to 0$ we have $[a_{\mu}]^{4G2HDM}_{H^{\pm}} \to 0$,
while
\be
F_A^H \left( \delta_{U_2} \to 0 \right) \approx \frac{i\,f_{B_s}}{4} \frac{S_{t's}^{H^-} {S_{t'b}^{H^-}}^{\ast}}{M_{H^-}^2}
B_{HH}^A(y_{t'},y_{\nu'})
\frac{m_{\tau^\prime}^2}{m_W^2} f_\beta^2 \left| U_{44} \right|^2
\left| \Sigma_{44}^\nu \right|^2 \left|\delta_{\Sigma_2} \right|^2 ~,
\ee
The contributions from the charged Higgs exchange diagrams with ($t,\nu'$) can be obtained directly from
Eqs.~\ref{FSH},\ref{FPH} and \ref{formfactors} by replacing $t'$ with $t$, which eventually requires
the replacements: $t' \leftrightarrow t$ and $\Sigma^u_{4 \leftrightarrow 3}$ in Eq.~\ref{qcoup}.
\begin{widetext}
\begin{figure}[htb]
\begin{center}
\epsfig{file=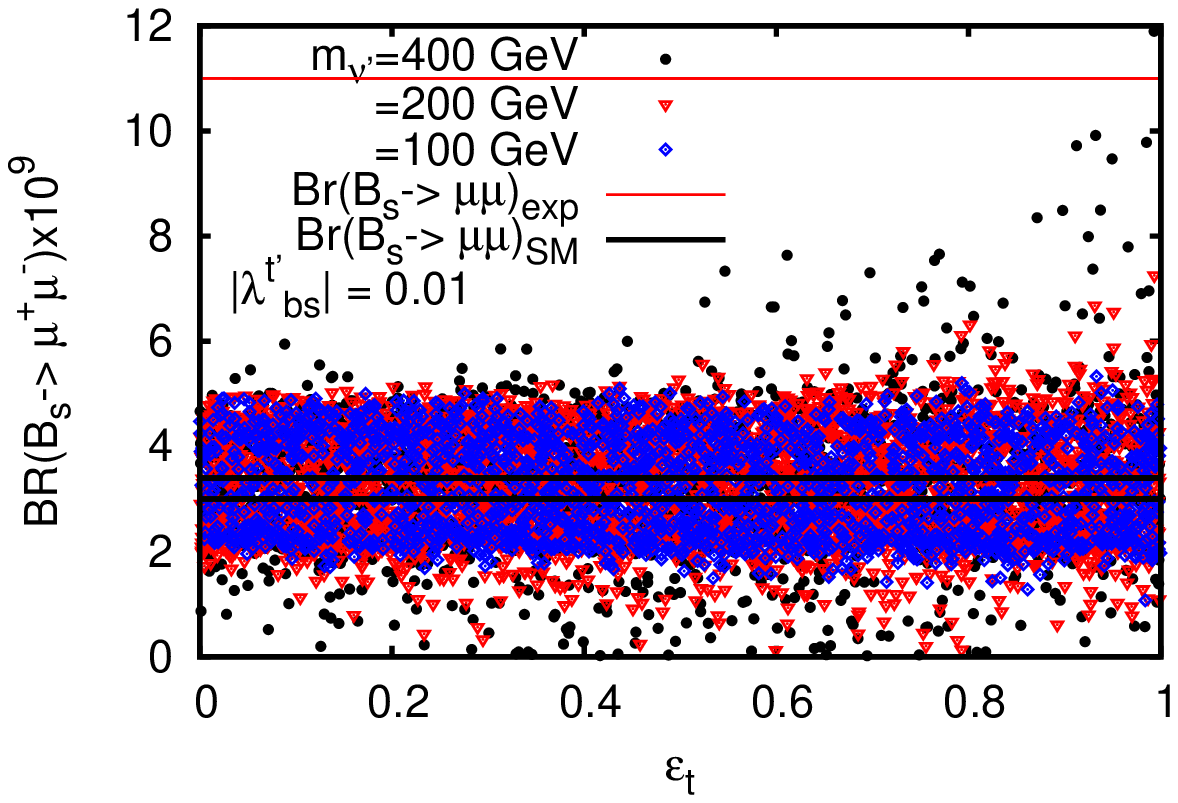,height=7cm,width=7cm,angle=0}
\epsfig{file=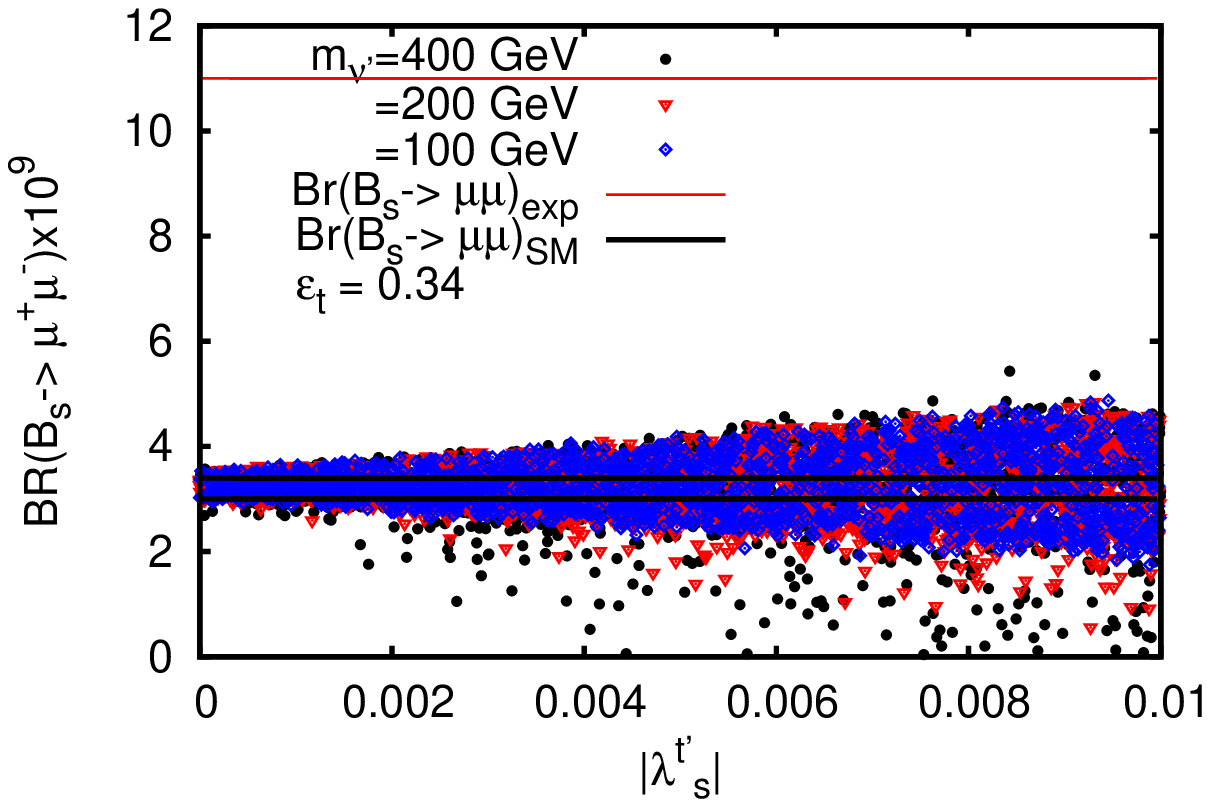,height=7cm,width=7cm,angle=0}
\caption{\emph{$BR(B_s \to \mu \mu)$ in the 4G2HDM
from box diagrams with the $H^+$ and $(t, \nu^\prime)$, $(t^\prime, \nu^\prime)$ exchanges,
as a function of $\epsilon_t$ (fixing $\epsilon_b=0.01$)
and $\lambda_{bs}^{t^\prime} \equiv V_{t^\prime b} V_{t^\prime s}^\ast$,
for $m_{\nu^\prime}=100,~200,~400$ GeV, $m_{\tau^\prime} = m_{\nu^\prime}$
and $m_{H^+}=500$ GeV. We considered only values of $\delta_{U_2}$ and $\delta_{\Sigma_2}$
which are allowed by $a_{\mu}$ given in Eq.~\ref{AMM}, keeping both of them $\lsim 0.2$.
Also shown are the experimental 95\% CL upper bound (upper/red horizontal line) and the SM predicted
range ($1 \sigma$) of values (lower/black horizontal lines).
}}
\label{bsmumu}
\end{center}
\end{figure}
\end{widetext}

In Fig.~\ref{bsmumu} we plot the contribution to $BR(B_s \to \mu \mu)$ in the 4G2HDM
from the box diagrams with the $H^+$ and $(t, \nu^\prime)$, $(t^\prime, \nu^\prime)$ exchanges,
as a function of $\epsilon_t$ (fixing $\epsilon_b=0.01$) and
$\lambda_{bs}^{t^\prime} \equiv V_{t^\prime b} V_{t^\prime s}^\ast$,
for $m_{\nu^\prime}=100,~200,~400$ GeV, $m_{\tau^\prime} = m_{\nu^\prime}$ and
$m_{H^+}=500$ GeV. The allowed ranges of the key parameters $\delta_{U_2}$ and $\delta_{\Sigma_2}$,
which control the muon $g-2$ in our model, are randomly chosen
in the range $[0,0.2]$ to be consistent with $a_{\mu}^{exp}$ given in Eq.~\ref{amuexp}.
We also show the current experimental bound \cite{lhcb1} and the SM predicted value for $BR(B_s \to \mu \mu)$.
We see that the contribution from the new (i.e., in the 4G2HDM) box diagrams that
involve the heavy 4th generation neutrino is consistent with the current experimental bound
on $BR(B_s \to \mu \mu)$ for values of $\delta_{U_2}$ and $\delta_{\Sigma_2}$ that
reproduce the observed muon $g-2$. It is also interesting to note that both in the SM4 and in the 4G2HDM,
Br$(B_s \to \mu^+\mu^-)$ can differ from the SM value by at-most a factor of \cal{O}(3) in either direction.

\section{Summary and discussion \label{sec5}}

We have considered the effects of 1-loop exchanges of heavy 4th generation
leptons on the muon $g-2$, on the lepton flavor violating decays $\mu \to e \gamma$,
$\tau \to \mu \gamma$ and on $B_s \to \mu^+ \mu^-$,
in the 4G2HDM which is a 2HDM where the Higgs doublet with the heavier VEV is coupled only to
the 4th generation doublet while the ``lighter" Higgs doublet is coupled to
fermions of the 1st-3rd generations. This model is particularly motivated
for the leptonic sector, as it effectively
addresses the
heaviness of a 4th generation EW-scale neutrino.

The muon $g-2$ is sensitive in our model to the product
$\delta_{U_2} \cdot \delta_{\Sigma_2}$, where
$\delta_{U_2} \equiv \frac{U_{24}^*}{U_{44}^*}$,
$\delta_{\Sigma_2} \equiv \frac{\Sigma_{42}^{e *}}{\Sigma_{44}^\nu}$,
$U_{ij}$ is the leptonic CKM-like PMNS matrix and
$\Sigma_{ij}^e,~\Sigma_{ij}^\nu$ are new mixing matrices in the charged
and neutral leptonic sectors that are unique to the 4G2HDM.

We find that, depending on the mass $m_{\nu'}$, the experimentally measured muon magnetic moment can
be accounted for if ${\cal O}(10^{-3}) \lsim \delta_{U_2} \cdot \delta_{\Sigma_2}
\lsim {\cal O}(10^{-2})$. We also find that the decays
$\mu \to e \gamma$ and
$\tau \to \mu \gamma$ can have branching ratios which are
not too far below the current bounds,
i.e., of ${\cal O}({\rm few} \cdot 10^{-13})$
and ${\cal O}({\rm few} \cdot 10^{-9})$, respectively,
if the products $\delta_{U_1} \cdot \delta_{\Sigma_2},~\delta_{U_2} \cdot \delta_{\Sigma_1} \sim {\cal O}(10^{-6})$
and $\delta_{U_2} \cdot \delta_{\Sigma_3},~\delta_{U_3} \cdot \delta_{\Sigma_2} \sim {\cal O}(10^{-3})$, respectively.

We also considered the effects of one-loop exchanges of the 4th generation heavy neutrino $\nu'$
on the decay $B_s \to \mu^+ \mu^-$ and found that,
in the four generations
model considered here, ${\rm BR}(B_s \to \mu^+ \mu^-)$ can be larger or smaller
than the SM predicted value by a factor of about three, for
values of
${\cal O}(10^{-3}) \lsim \delta_{U_2} \cdot \delta_{\Sigma_2}
\lsim {\cal O}(10^{-2})$, which render the observed value of the muon
$g-2$ to be
consistent with the current upper limit on this decay.

\bigskip

{\bf Acknowledgments:} SBS acknowledges research support from the Technion.
The work of AS was supported in part by the U.S. DOE contract
\#DE-AC02-98CH10886(BNL).

\end{document}